\documentclass[twocolumn,iop,apj]{emulateapj}
\usepackage{times}
\usepackage{natbib}
\usepackage[draft]{hyperref}

\usepackage{verbatim}
\usepackage[normalem]{ulem}
\usepackage[usenames, dvipsnames]{color}
\newcommand{\imos}[1]{#1} 

\newcommand{\myemail}{Corresponding mail}

\newcommand{\fermilat}{\emph{Fermi}-LAT}

\newcommand{\galprop}{GALPROP}
\newcommand{\gray}{$\gamma$-ray}

\newcommand{\helmod}{\textsc{HelMod}}
\newcommand{\rigidity}{R}

\slugcomment{Submitted to ApJ}

\shorttitle{Local interstellar spectra of primary nuclei}
\shortauthors{Boschini et al.}

\usepackage{amsmath} 

\begin{document}


\title{Deciphering the local Interstellar spectra of primary cosmic ray species with \helmod{}
}


\author{
M.~J.~Boschini\altaffilmark{1,2},  
S.~{Della~Torre}\altaffilmark{1}, 
M.~Gervasi\altaffilmark{1,3}, 
D.~Grandi\altaffilmark{1},
G.~J\'{o}hannesson\altaffilmark{4,5}, 
G.~{La~Vacca}\altaffilmark{1}, 
N.~Masi\altaffilmark{6},
I.~V.~Moskalenko\altaffilmark{7,8}, 
S.~Pensotti\altaffilmark{1,3}, 
T.~A.~Porter\altaffilmark{7,8}, 
L.~Quadrani\altaffilmark{6,9}, 
P.~G.~Rancoita\altaffilmark{1},
D.~Rozza\altaffilmark{1,3}, 
and M.~Tacconi\altaffilmark{1,3}
}
\email{\myemail}


\altaffiltext{1}{INFN, Milano-Bicocca, Milano, Italy}
\altaffiltext{2}{also CINECA, Segrate, Milano, Italy}
\altaffiltext{3}{also Physics Department, University of Milano-Bicocca, Milano, Italy}
\altaffiltext{4}{Science Institute, University of Iceland, Dunhaga 3, IS-107 Reykjavik, Iceland}
\altaffiltext{5}{also NORDITA,  Roslagstullsbacken 23, 106 91 Stockholm, Sweden}
\altaffiltext{6}{INFN, Bologna, Italy}
\altaffiltext{7}{Hansen Experimental Physics Laboratory, Stanford University, Stanford, CA 94305}
\altaffiltext{8}{Kavli Institute for Particle Astrophysics and Cosmology, Stanford University, Stanford, CA 94305}
\altaffiltext{9}{also, Physics Department, University of Bologna, Bologna, Italy}


\begin{abstract}

Local interstellar spectra (LIS) of primary cosmic ray (CR) nuclei, such as helium, oxygen, and mostly primary carbon are derived for the rigidity range from 10 MV to $\sim$200 TV using the most recent experimental results combined with the state-of-the-art models for CR propagation in the Galaxy and in the heliosphere. Two propagation packages, \galprop{} and \helmod{}, are combined into a single framework that is used to reproduce direct measurements of CR species at different modulation levels, and at both polarities of the solar magnetic field. The developed iterative maximum-likelihood method uses \galprop-predicted LIS as input to \helmod{}, which provides the modulated spectra for specific time periods of the selected experiments for model-data comparison. The interstellar and heliospheric propagation parameters derived in this study are consistent with our prior analyses using the same methodology for propagation of CR protons, helium, antiprotons, and electrons. The resulting LIS accommodate a variety of measurements made in the local interstellar space (Voyager 1) and deep inside the heliosphere at low (ACE/CRIS, HEAO-3) and high energies (PAMELA, AMS-02). 
\end{abstract}



\section{Introduction} \label{Introduction} \label{Intr}

New instrumentation launched into space over the last decade (PAMELA, \citealt{2007APh....27..296P}, {\it Fermi} Large Area Telescope, \citealt{2009ApJ...697.1071A}, AMS--02, \citealt{2013PhRvL.110n1102A}) signify the beginning of a new era in astrophysics. New technologies employed by these space missions have enabled measurements with unmatched precision, which allows for searches of subtle signatures of new phenomena in CR and \gray{} data. Combined with the results of past missions, such as ATIC, BESS, CAPRICE, CREAM, HEAO-3, HEAT, ISOMAX, TIGER and SuperTIGER, TRACER, Ulysses, and those that are still running, such as Voyager~1 and ACE/CRIS, this led to a remarkable progress in the field of astrophysics of CRs that was established more than one hundred years ago. Unique in these series of experiments are the Voyager~1,~2 spacecrafts launched in 1977 and whose on-board instruments are providing data on the elemental spectra and composition at the interstellar reaches of the Solar system \citep{2013Sci...341..150S,2016ApJ...831...18C}.

Other high-expectations missions that started to deliver breakthrough data are CALET, DAMPE, and ISS-CREAM. Indirect CR measurements are made through observations of their emissions by space- and ground-based telescopes: INTEGRAL, HAWC, H.E.S.S., MAGIC, VERITAS, WMAP, and Planck. The most spectacular is the \fermilat{} mission that is mapping the all-sky diffuse \gray{} emission produced by CR interactions in the interstellar medium (ISM) and in the vicinity of CR accelerators. 

Primary nuclei, such as helium, oxygen, and mostly primary carbon, are the most abundant species in CRs after hydrogen and are the priority targets for CR missions. They are also the most abundant in the Universe, thanks to the primordial nucleosynthesis of helium and stellar nucleosynthesis that provides the heavier species. Their fragmentation in CRs produces the majority of lighter nuclides and is the main source of lithium, beryllium, and boron, which are termed ``secondary.'' The ratios of secondary-to-primary species in CRs can be used to study properties of CR propagation in the Galaxy and provide a basis for other related studies, such as processes of particle acceleration, CR sources, properties of the ISM, search for signatures of new physics, and many others. Therefore, the LIS of helium, carbon, and oxygen are of considerable interest for astrophysics and particle physics. 

Recently we demonstrated that by combining two packages, \galprop{} for interstellar propagation, and \helmod{} for heliospheric propagation, into a single framework we were able to reproduce the direct measurements of CR protons, helium, antiprotons, and electrons \citep{2017ApJ...840..115B,2018ApJElectron} made by Voyager 1, BESS, PAMELA, AMS-01, and AMS-02 at different modulation levels, and at both polarities of the solar magnetic field. The employed iterative method uses \galprop-predicted LIS as input to \helmod{}, which provides the modulated spectra for specific time periods of the selected experiments for model-data comparison. The derived LIS of CR species can be used to facilitate significantly studies of CR propagation in the Galaxy and in the heliosphere by disentangling these two massive tasks and will lead to further progress in understanding of both processes. In this paper we extend this approach to derive the LIS of carbon and oxygen over a wide range of rigidities from 10 MV to $\sim$200 
TV using the most recent data \citep{2017PhRvL.119}. The helium LIS obtained from our earlier analysis is also re-evaluated using the updated results from AMS-02. 


\section{CR transport in the Galaxy and the heliosphere} \label{sec2}

\subsection{\galprop{} Model for Galactic CR Propagation and diffuse emission}
\label{galprop}

Understanding of the origin of CRs, their acceleration mechanisms, main features of their interstellar propagation, and the CR source composition requires both precise observational data and a strong theoretical effort \citep{2007ARNPS..57..285S}. A unification of many different kinds of data into a self-consistent picture requires a state-of-the-art numerical tool that incorporates the latest information on the Galactic structure (distributions of gas, dust, radiation and magnetic fields), the up-to-date formalisms describing particle and nuclear cross sections, and a full theoretical description of the processes in the ISM.  This was realized about 20 years ago, when some of us started to develop the most advanced fully numerical CR propagation code, called \galprop{}\footnote{Available from http://galprop.stanford.edu \label{galprop-site}} \citep{1998ApJ...493..694M,1998ApJ...509..212S}. 

The key idea behind  \galprop{} is that all CR-related data, including direct measurements, \gray{s}, synchrotron radiation, etc., are subject to the same Galactic physics and must be modeled simultaneously. Since the beginning of the project, the \galprop{} model for CR propagation is being continuously developed in order to provide a framework for studies of CR propagation in the Galaxy and interpretation of relevant observations \citep{2007ARNPS..57..285S,1998ApJ...493..694M,1998ApJ...509..212S,2000ApJ...528..357M,2000ApJ...537..763S,2004ApJ...613..962S,2002ApJ...565..280M,2003ApJ...586.1050M,2006ApJ...642..902P,2011ApJ...729..106T,2011CoPhC.182.1156V,2012ApJ...752...68V,2016ApJ...824...16J}. The latest version and supplementary datasets are available through a WebRun interface at the dedicated website\textsuperscript{\ref{galprop-site}}. 

In this work we use a newly developed version 56 of the \galprop{} code, which is described in \citet{PoS(ICRC2015)492} and \citet{2017ApJ...846...67P}, and references therein. The current version has the ability to assign the injection spectrum independently to each isotope. It also builds a dependency tree for the isotopes included in each run from the nuclear reaction network to ensure that dependencies are propagated before the source term is calculated. This way, special cases of $\beta^-$-decay (e.g., $^{10}$Be$\to^{10}$B) are treated properly in one pass of the reaction network, instead of the two passes required before, thus providing a significant gain in speed.

The procedure of intercalibration between \helmod{} and \galprop{}, described by \citet{2017ApJ...840..115B}, uses proton spectra as a reference for evaluating the modulation parameters assuming that all Galactic CRs species are subject to the same heliospheric conditions in the considered energy range.  The resulting \helmod{} and \galprop{} set of parameters was applied directly to CR LIS derived from the \galprop{} MCMC scan and compared with the direct measurements at 1 au (see Section~\ref{Sect::DataAtEarth}).

\subsection{\helmod{} Model for heliospheric transport}
\label{Sect::Helmod}

CR spectra computed by \galprop{} cannot be directly compared with low-energy CR observations made deep inside the heliosphere (typically at Earth's orbit) due to specific properties of the interplanetary medium that have to be addressed separately~\citep[see discussion in][]{2017ApJ...840..115B}.  CR propagation in the heliosphere was first studied by \citet{1965P&SS...13....9P}, who formulated the transport equation, also referred to as the Parker equation~\citep[see, e.g., discussion in][and reference therein]{Bobik2011ApJ}:
\begin{align}
\label{EQ::FPE}
 \frac{\partial U}{\partial t}= &\frac{\partial}{\partial x_i} \left( K^S_{ij}\frac{\partial \mathrm{U} }{\partial x_j}\right)\\
&+\frac{1}{3}\frac{\partial V_{ \mathrm{sw},i} }{\partial x_i} \frac{\partial }{\partial T}\left(\alpha_{\mathrm{rel} }T\mathrm{U} \right)
- \frac{\partial}{\partial x_i} [ (V_{ \mathrm{sw},i}+v_{d,i})\mathrm{U}],\nonumber
\end{align}
where $U$ is the number density of Galactic CR particles per unit of kinetic energy $T$, $t$ is time, $V_{ \mathrm{sw},i}$ is the solar wind velocity along the axis $x_i$, $K^S_{ij}$ is the symmetric part of the diffusion tensor, $v_{d,i}$ is the particle magnetic drift velocity (related to the antisymmetric part of the diffusion tensor), and finally $\alpha_{\mathrm{rel} }=\frac{T+2m_r c^2}{T+m_r c^2} $, with $m_r$ the particle rest mass in units of GeV/nucleon. Parker's transport-equation describes:
i) the \textit{diffusion} of Galactic CRs by magnetic irregularities,
ii) the so-called \textit{adiabatic-energy changes} associated with expansions and compressions of cosmic radiation,
iii) an \textit{effective convection} resulting from the convection with the \textit{solar wind} (SW, with velocity $\vec{V}_{{\rm sw}}$), and iv) the drift effects related to the \textit{drift velocity} ($\vec{v}_d$).

The overall effect of heliospheric propagation on Galactic CRs is a reduction of measured CR intensities that depends on the solar activity and is called ``solar modulation.'' In this work, the particle transport within the heliosphere, from the Termination Shock (TS) down to Earth orbit, is described using the \helmod{} code\footnote{http://www.helmod.org/} \citep[and reference therein]{2017HelMod}. The \helmod{} code integrates the \citet{1965P&SS...13....9P}  transport equation using a Monte Carlo approach involving stochastic differential equations~\citep[see a discussion in, e.g.,][]{Bobik2011ApJ,BobikEtAl2016}. 

\imos{The present form of the diffusion parameter \citep[as defined in][and reference therein]{2017HelMod} includes a scale correction factor that rescales the absolute value proportionally to the drift contribution. As discussed by \citet{2017ApJ...840..115B} this correction is evaluated for proton spectra during the positive HMF polarity period to account for the presence of the latitudinal structure in the spatial distribution of Galactic CRs. The effect of turbulence on the diffusion term is accounted by the value of the $g_{low}$ term in the expression for $K_{||}$ \citep[see Section 3 of][]{2017ApJ...840..115B}. }

\imos{The presence of turbulence in the interplanetary medium should reduce the global effect of CR drift in the heliosphere \citep[see a discussion in Sections 2 and 3 of][]{2017ApJ...840..115B}, and this is usually accounted by a drift suppression factor that is more relevant at rigidities below 1 GV. As also discussed by \citet{2017HelMod}, the validity of the \helmod{} code is verified down to 1 GV in rigidity. Lower energies would require further improvement in the description of the solar modulation in the outer heliosphere \citep[see, e.g.,][]{2011ApJ...735..128S,2017NatAs...1E.115D} -- from the TS up to the interstellar space -- as well as inclusion of turbulence in the evaluation of the drift term \citep[see, e.g.,][]{2017ApJ...841..107E}. Such an additional treatment may have an impact on modulated spectra at low energies during the periods of intermediate activity. However, detailed investigation of this effect requires observational data with high statistics and correspondingly small statistical and systematic errors.}

\begin{figure}
\begin{center}
 \includegraphics[width=0.49\textwidth]{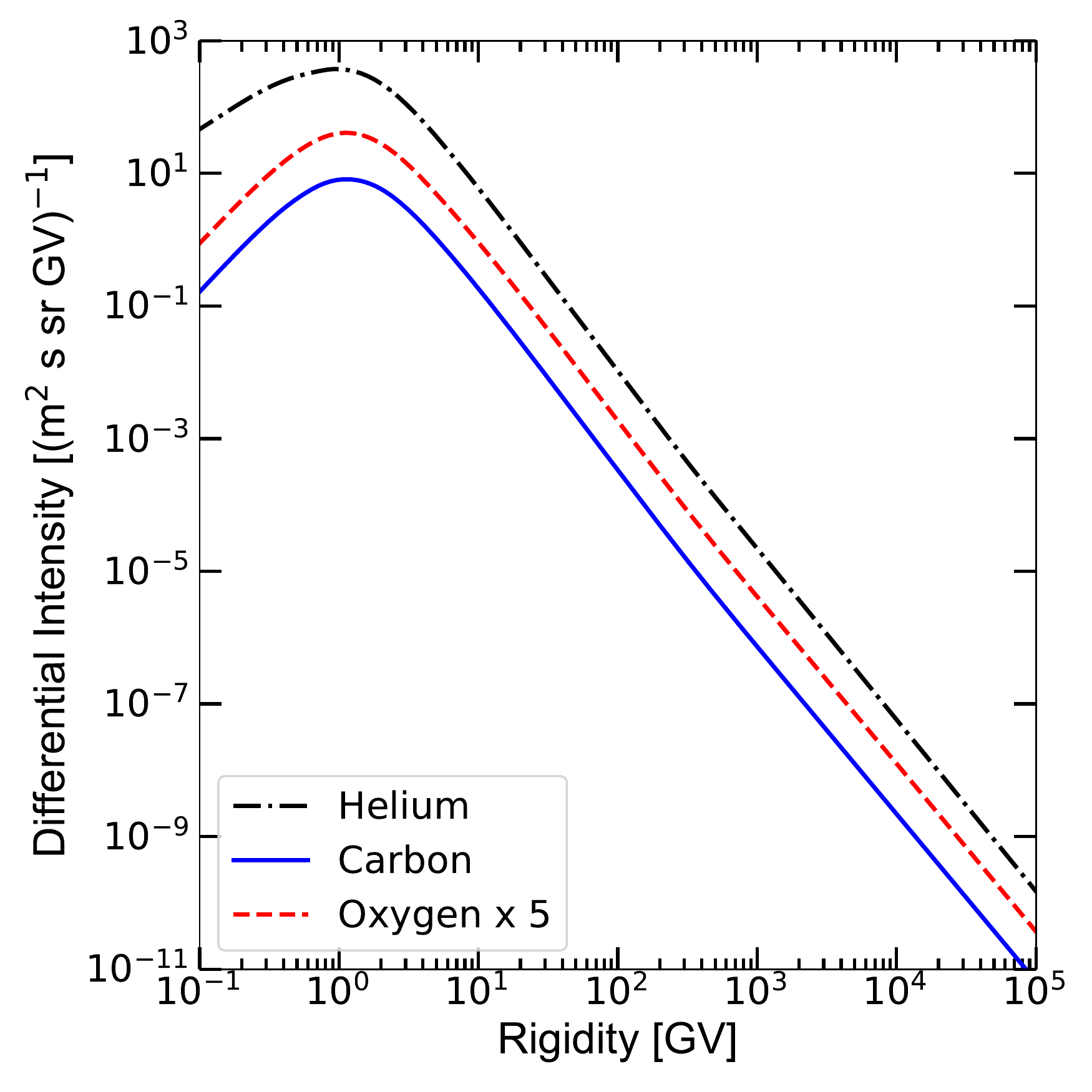}
 \caption{The local interstellar spectra of CR helium (black dash-dotted line), carbon (blue solid line), and oxygen (red dashed line) as derived from the MCMC procedure (see text). Oxygen LIS is multiplied by a factor of 5 to avoid overlapping with carbon LIS.
 \label{fig:GALPROPLISS}}
 \end{center}
\end{figure}

\section{Interstellar propagation}

The LIS of CR species shown in Figure~\ref{fig:GALPROPLISS} are obtained using the optimization procedure employing \galprop{} and \helmod{} codes in concert. The combined framework, described by \citet{2017ApJ...840..115B}, is logically divided into two parts:  i) a MCMC interface to version 56 of \galprop{}~\citep{MasiDM2016}, that enables the sampling of the CR production and propagation parameters space, and ii) an iterative procedure that, starting from \galprop{} output, provides modulated spectra computed with \helmod{} to compare with AMS-02 data as observational constraints \citep{2017HelMod}. The final product is a set of Galactic and heliospheric propagation parameters for all CR species that provides LIS that best reproduce available experimental data.

The basic features of CR propagation in the Galaxy are well-known, but the exact values of propagation parameters depend on the assumed propagation model and accuracy of selected CR datasets. Therefore, the MCMC procedure is employed to determine the propagation parameters using the best available CR measurements. Five main propagation parameters, that affect the overall shape of CR spectra, were left free in the scan using \galprop{} running in the 2D mode: the Galactic halo half-width $z_h$, the normalization of the diffusion coefficient $D_0$ and the index of its rigidity dependence $\delta$, the Alfv\'en velocity $V_{\rm Alf}$, and the gradient of the convection velocity $dV_{\rm conv}/dz$ ($V_{\rm conv}=0$ in the plane, $z=0$). The radial size of the Galaxy does not significantly affect the values of propagation parameters and was set to 20 kpc. In addition, a factor $\beta^\eta$ is included in the diffusion coefficient, where $\beta=v/c$ and $\eta$ was left free. The best-fit value of $\eta=0.71$ improves the agreement at low energies, and slightly affects the choice of injection indices ${\gamma}_0$ an ${\gamma}_1$ (Table~\ref{tbl-2}).

Inclusion of both distributed reacceleration and convection simultaneously is necessary to describe the high precision AMS-02 data, particularly in the range below 10 GV where their effects on CR spectra are significant \citep[see][ for more details]{2017ApJ...840..115B}. The best-fit values of the main propagation parameters tuned to AMS-02 data are listed in Table~\ref{tbl-1}, which are similar to those obtained by \citet{2017ApJ...840..115B} within the quoted error bars. The most significant change is a slight increase of the Alfv\'en velocity $V_{\rm Alf}$ by 2.4 km s$^{-1}$ which improves the agreement with the B/C ratio and electron data \citep{2018ApJElectron}.

The MCMC procedure is used only in the first step to define a consistent parameter space, then a methodical calibration of the model employing the \helmod{} module is performed. \imos{This procedure is described in detail in Section 3 of \citet{2017ApJ...840..115B}, pp.\ 7, 8. Parameters of the injection spectra at low energies, such as spectral indices $\gamma_i (i=0,1,2)$ and the break rigidities $R_i (i=0,1)$,}
are left free in the procedure at this stage because their exact values depend on the solar modulation. Hence the low energy parts of the injection spectra are tuned together with the solar modulation parameters \imos{within their physical ranges in order to find best-fit solutions for all the observables. Consequently, the final values are coming from the \galprop-\helmod{} combined fine-tuning, which involves an exploration of the parameter space around the best values defined in the first step.}

\begin{deluxetable}{crlc}[!tb]
\tablecolumns{4}
\tablewidth{0mm}
\tablecaption{Best-fit propagation parameters \label{tbl-1}}
\tablehead{
\colhead{N} &
\multicolumn{2}{c}{Parameter\quad} &
\colhead{Best Value} 
}
\startdata
1 & $z_h$,& kpc &$4.0\pm0.7$ \\
2 & $D_{0}$,& $10^{28}$ cm$^{2}$ s$^{-1}$  &$4.3\pm0.6$ \\
3 & $\delta$ &&$0.415\pm0.025$\\
4 & $V_{\rm Alf}$,& km s$^{-1}$ &$31\pm3$\\
5 & $dV_{\rm conv}/dz$,& km s$^{-1}$ kpc$^{-1}$ & $9.8\pm 0.7$
\enddata
\end{deluxetable}

\begin{deluxetable}{crrrrr}[!tb]
\tablecolumns{6}
\tablecaption{Spectral parameters for CR species \label{tbl-2} }
\tablehead{
\colhead{Parameters} &
\multicolumn{1}{c}{He} &
\multicolumn{1}{c}{C} &
\multicolumn{1}{c}{O} &
\colhead{Error} &
\colhead{$s_i$}
}
\startdata
$R_{0}$ &\nodata &1 GV & 1 GV &0.5 & $-0.15$ \\
$R_{1}$ &7 GV &7 GV & 7 GV &1& $-0.20$ \\
$R_{2}$ &325 GV &345 GV &360 GV &15 & 0.15 \\
${\gamma}_0$ & \nodata &1 &1.1 &0.06 & \nodata \\
${\gamma}_1$ &1.76 &1.98 &1.99 &0.06& \nodata \\
${\gamma}_2$ &2.39 &2.42 &2.46 &0.04& \nodata \\
${\gamma}_3$ &2.15 &2.12 &2.13 &0.04& \nodata 
\enddata
\end{deluxetable}

\begin{figure*}[!tb]
\centerline{
\includegraphics[width=0.8\textwidth]{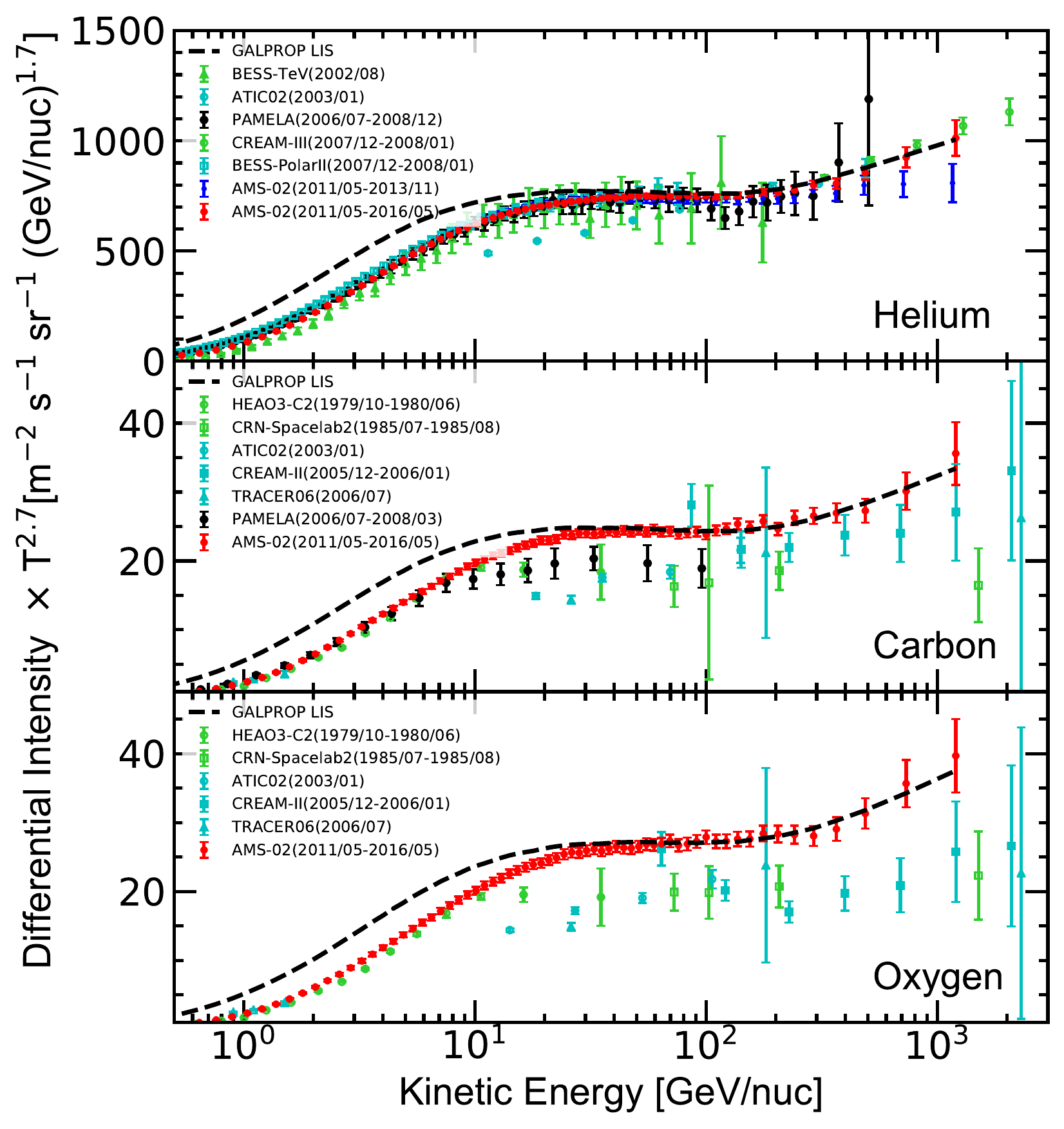}
}
\caption{The best fit helium LIS (top panel), carbon LIS (middle panel), and oxygen LIS (bottom panel) as a function of kinetic energy per nucleon shown together with the AMS-02 data \citep[2011/05-2016/05,][]{2017PhRvL.119}, and previous measurements by ATIC \citep{2009BRASP..73..564P}, BESS \citep{bess_prot,BESS2007_Abe_2016}, CREAM \citep{2009ApJ...707..593A,2017ApJ...839....5Y}, HEAO3-C2 \citep{1990AeA...233...96E}, CRN-Spacelab2 \citep{1991ApJ...374..356M}, TRACER \citep{2011ApJ...742...14O}, and PAMELA \citep{2011Sci...332...69A,2014ApJ...791...93A}. Data points are taken from the LPSC Database of Charged Cosmic Rays \citep{2014A&A...569A..32M}, and the corresponding experiments are listed in Table~\ref{tbl-3}.}
\label{fig:HighEnergy_He_C_O}
\end{figure*}

\begin{deluxetable*}{lcl|lcl}[!tb]
\tablecolumns{6}
\tablecaption{A list of high energy data used in this paper \label{tbl-3}}
\tablehead{
\colhead{Experiment} &
\colhead{CR species} & 
\colhead{Reference}  &
\colhead{Experiment} &
\colhead{CR species} & 
\colhead{Reference}  
}
\startdata
HEAO3-C2 (1979/10-1980/06)  &C, O   & \citet{1990AeA...233...96E}       &PAMELA (2006/07-2008/03)    &C  &\citet{2014ApJ...791...93A}            \\
CRN-Spacelab2 (1985/07-1985/08) &C, O   & \citet{1991ApJ...374..356M}   &PAMELA (2006/07-2008/12)    &He& \citet{2011Sci...332...69A}            \\ 
BESS-TeV (2002/08)          &He & \citet{bess_prot}                     & CREAM-III (2007/12-2008/01)     &He& \citet{2017ApJ...839....5Y}        \\
ATIC02 (2003/01)        &He, C, O   & \citet{2009BRASP..73..564P}       &BESS-PolarII (2007/12-2008/01)  &He& \citet{BESS2007_Abe_2016}          \\
CREAM-II (2005/12-2006/01)  &C, O   &\citet{2009ApJ...707..593A}        &AMS-02 (2011/05-2013/11)    &He& \citet{2015PhRvL.115u1101A}            \\ 
TRACER06 (2006/07)      &C, O   &\citet{2011ApJ...742...14O}            &AMS-02 (2011/05-2016/05)        &He, C, O& \citet{2017PhRvL.119}
\enddata
\end{deluxetable*}

\imos{In order to avoid sharp unphysical breaks in the LIS, we use the following parameterization for the source injection spectrum:
\begin{equation}
q(\rigidity) \propto \left(R/\rigidity_{0}\right)^{-\gamma_0} \prod_{i=0}^{2} \left[1 + \left(R/\rigidity_{i}\right)^{\frac{\gamma_i-\gamma_{i+1}}{s_i} }\right]^{s_i},
\label{injection}
\end{equation}
where $s_i$ are the smoothing parameters, $s_i\lessgtr0$ if $|\gamma_i |\lessgtr |\gamma_{i+1} |$. The values of $s_i (i=0,1,2)$ used in the fits are provided in Table \ref{tbl-2}.}

Reproduction of the spectra of primary nuclei from MeV to TeV energies altogether requires an injection spectrum with three breaks. The MCMC scans in $\gamma_i$ and $R_i$ used the AMS-02 \citep{2017PhRvL.119} and Voyager~1 \citep{2016ApJ...831...18C} data as constraints. At the following step, these parameters were slightly modified together with parameters of the solar modulation in order to find the best-fit solution for the LIS, as explained by \citet{2017ApJ...840..115B}. Reproduction of the low-energy LIS corresponding to the direct measurements by Voyager 1 requires a break $R_{0}$ at 1 GV. The resulting best-fit spectral parameters are shown in Table~\ref{tbl-2}. 

The present analysis extends the dataset used in \citet{2017ApJ...840..115B} by including the newest published AMS-02 spectra of helium, carbon and oxygen that are based on the data collected during its first 5 years of operations~\citep{2017PhRvL.119}. These measurements represent the current \textit{state-of-art} knowledge of spectra of CR nuclei at high energies\footnote{The actual uncertainties of CR fluxes is actually far below the uncertainties related to, e.g., nuclear cross sections. Thus, in the MCMC scans we set a minimum errors of 5\% on the CR data, such as the B/C ratio, to allow a reasonable convergence of the Monte Carlo fitting procedure.}. In particular, it has been shown that the measured He-C-O spectra have a very similar rigidity dependence above 60 GV. They all deviate from a single power law above 200 GV and exhibit the same degree of spectral hardening \citep{2017PhRvL.119}. This spectral hardening is discussed in details in Section~\ref{Sect::highenergy}. The updated set of helium data (Figure~\ref{fig:HighEnergy_He_C_O}, top panel) motivates the fine tuning of the injection spectral index $\gamma_3$ above the high-energy break with respect to the previous analysis of helium data that was based on the three years AMS-02 dataset. The new value is listed in Table~\ref{tbl-2}. The helium LIS below 325 GV remains unchanged, \imos{and so do the modulated spectra,} and is the same as published by \citet{2017ApJ...840..115B}, therefore, it is not discussed further in this paper.

\begin{figure*}[thb]
\centerline{
\includegraphics[width=0.99\textwidth]{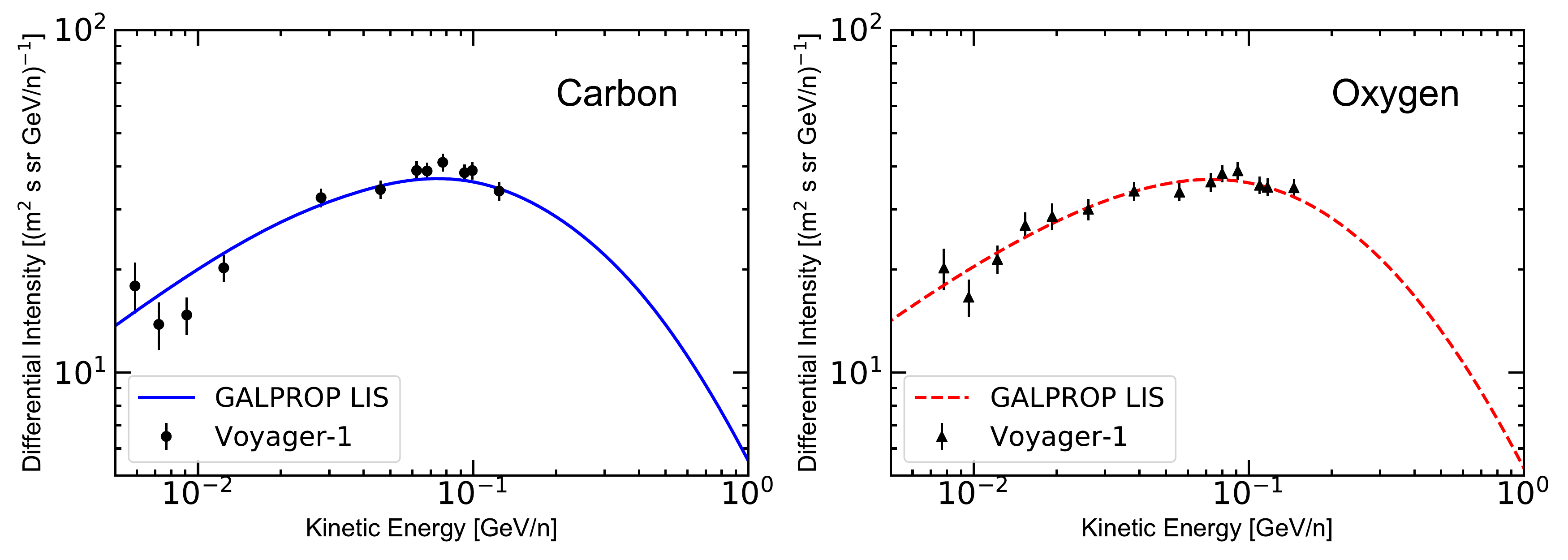}
}
\caption{A comparison of our best-set LIS (lines) with Voyager 1 2012-2015 monthly averaged data (points) shown as a function of kinetic energy per nucleon. Left panel: carbon spectrum, right panel: oxygen spectrum.}
\label{fig:Voyager_C_O}
\end{figure*}

\subsection{LIS of primary nuclei -- direct observations}\label{Sect::LISOutsideMod}

Accurate direct measurements of LIS are now available for several species at both low and high energies, and more is expected in coming years. Spectra of CR species above 100 GV are not affected by the heliospheric modulation and their measurements provide a direct probe of the LIS as discussed in details in Section~\ref{Sect::highenergy}. At low energies, from the second half of 2012, the Voyager 1 probe has been exploring the heliospheric boundaries providing invaluable data on the spectra of CR species in this region \citep{2013Sci...341..150S,2016ApJ...831...18C}. It is commonly accepted that Voyager 1 has been measuring the low-energy Galactic CRs from the local interstellar medium since that time \citep{2013Sci...341..144K,2014GeoRL..41.5325G}. 

The low-energy parameters of the injection spectra $\gamma_0$ and $R_0$ in the \galprop{} runs were tuned using Voyager 1 data as constraints for the MCMC scans. Moreover, their fine tuning was performed simultaneously with fine tuning of the solar modulation parameters (through the procedure described in Section~\ref{Sect::Helmod}) in order to find the best-fit solution, using the same procedure as described by \citet{2017ApJ...840..115B}. In Figure~\ref{fig:Voyager_C_O} the carbon and oxygen LIS are compared with their average intensities measured from the end of 2012 to the middle of 2015~\citep{2016ApJ...831...18C}. The presented model provides a good description of the LIS at low energies. The resulting best-fit spectral parameters are listed in Table~\ref{tbl-2} and remain the baseline solution for all scenarios discussed in this paper.

\begin{figure*}[tbp]
\centerline{
 \includegraphics[width=0.49\textwidth,height=0.43\textheight]{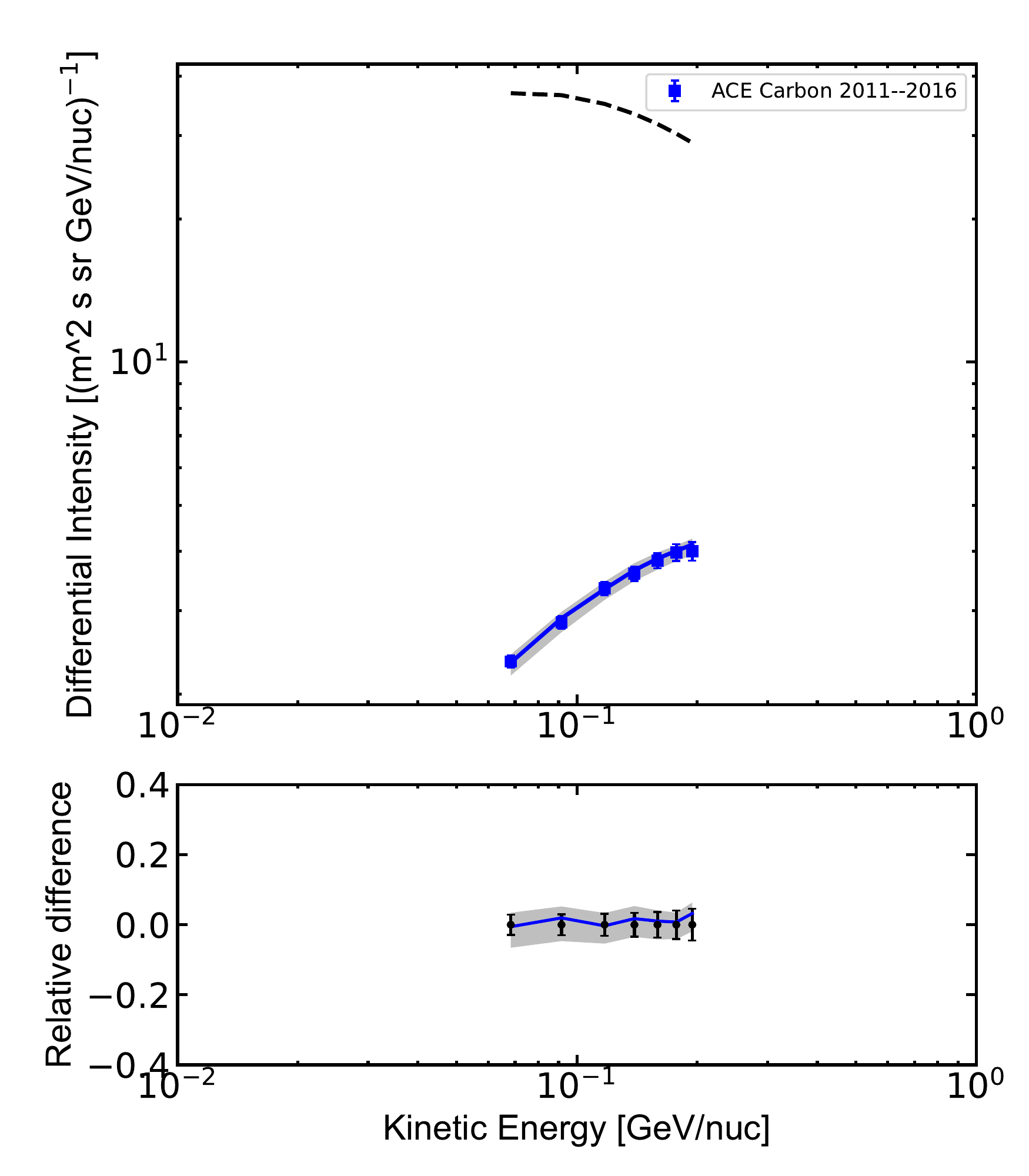}\hfill
 \includegraphics[width=0.49\textwidth,height=0.415\textheight]{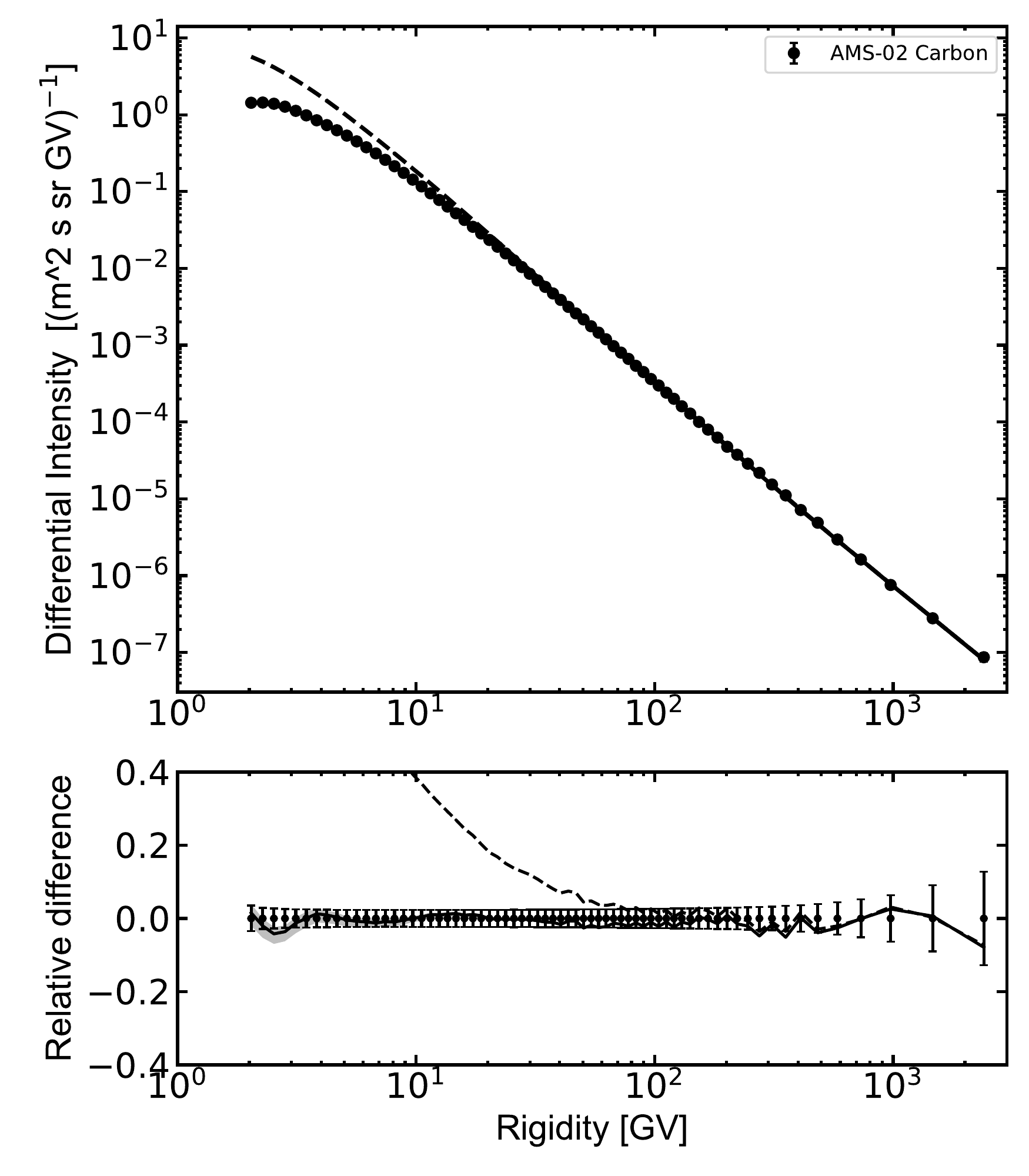}
 }
 \caption{Differential intensity of CR carbon, left panel: ACE/CRIS, right panel: AMS-02. Points represent experimental data, dashed line is the \galprop{} LIS, and solid line shows the computed modulated spectrum. The bottom panels show the relative difference between the modulated spectrum and experimental data. Data are presented in units of kinetic energy per nucleon or rigidity -- dependently on the experimental technique.
}
 \label{fig:C_AMS}
\end{figure*}

\begin{figure*}[tbp]
\centerline{
 \includegraphics[width=0.49\textwidth,height=0.43\textheight]{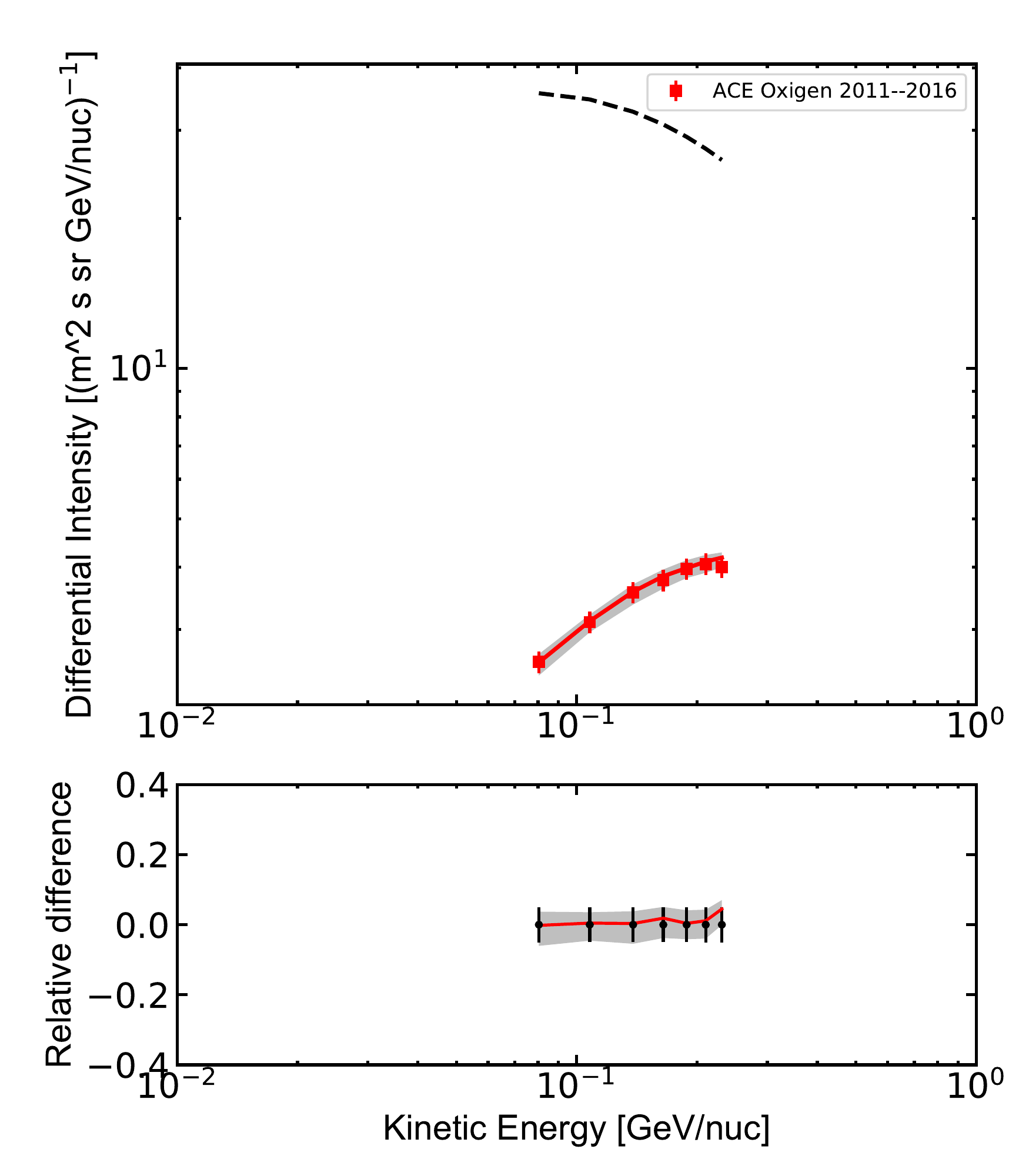}\hfill
 \includegraphics[width=0.49\textwidth,height=0.415\textheight]{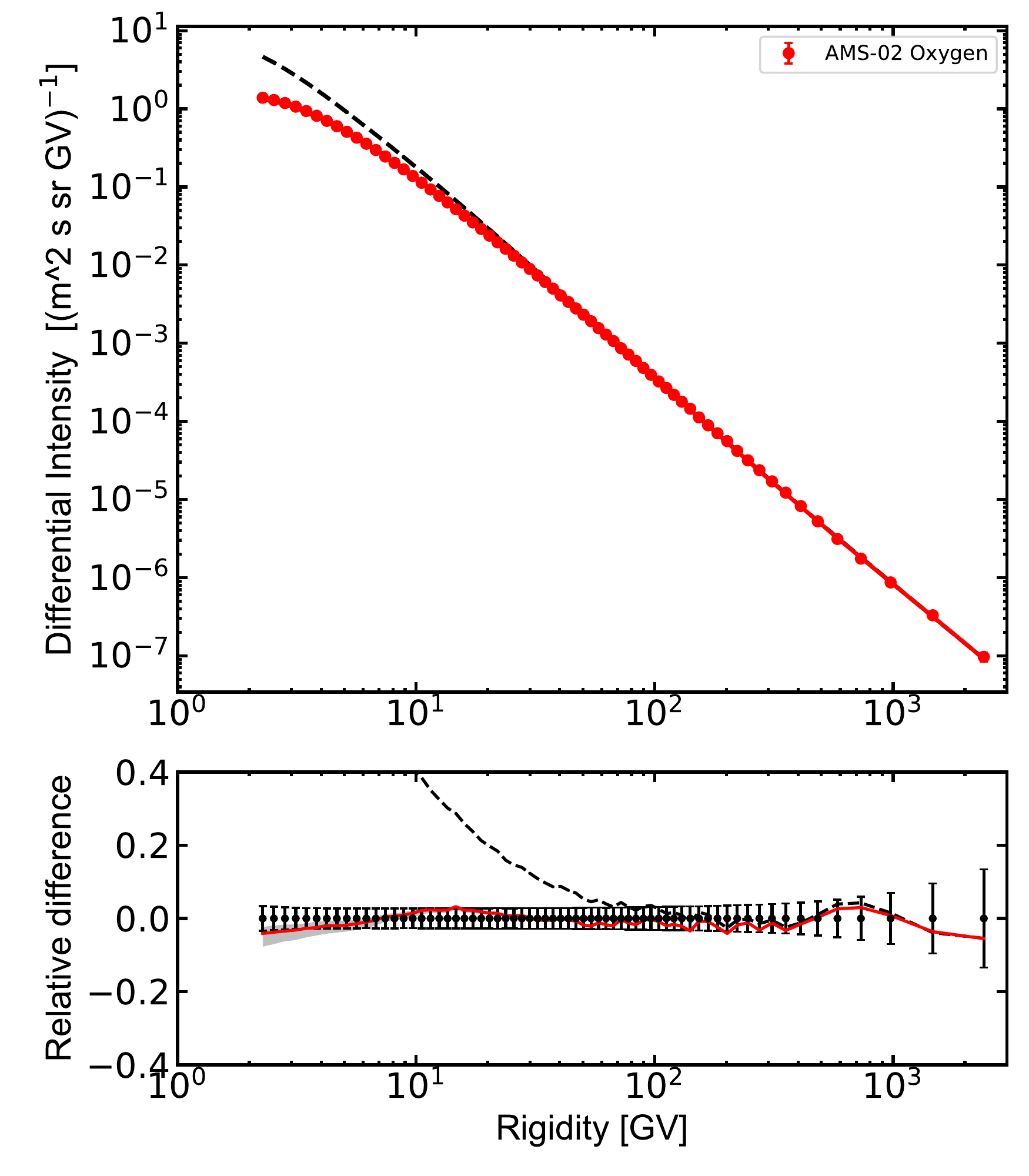}
 }
 \caption{Differential intensity of CR oxygen, left panel: ACE/CRIS, right panel: AMS-02. The bottom panels show the relative difference between the modulated spectrum and experimental data. Line coding is the same as in  Figure~\ref{fig:C_AMS}.
}
 \label{fig:O_AMS}
\end{figure*}

At high energies, where the CR spectra are not affected by the heliospheric modulation, AMS-02 data are used up to $\sim$2 TV, and further extension of the rigidity range to 20--30 TV is possible using high energy measurements (Table~\ref{tbl-3}) available from the LPSC Database of Charged Cosmic Rays~\citep{2014A&A...569A..32M}. More details are given in Section~\ref{Sect::highenergy}.

\subsection{The modulated spectrum: Data at Earth}\label{Sect::DataAtEarth}

Direct CR measurements that are made deep in the heliosphere, e.g., near the Earth, are affected by the solar modulation and cannot be compared directly to the LIS. Therefore, the \helmod{} code \citep{2017ApJ...840..115B} is used to calculate the modulated spectra of CR carbon and oxygen. The \helmod{} code, described in Section~\ref{Sect::Helmod}, was tuned to provide the same level of accuracy at any level of solar  activity, high and low, and for all kinds of isotopes ($Z/A$) propagating in the heliosphere using the same set of basic parameters \citep[see][for more details]{2017HelMod}. 

In this paper the modulated spectra of carbon and oxygen are compared with appropriate measurements by ACE/CRIS\footnote{http://www.srl.caltech.edu/ACE/ASC/} \citep{2009ApJ...698.1666G}, AMS-02 \citep{2017PhRvL.119},  HEAO-3~\citep{1990AeA...233...96E}, and PAMELA~\citep{2014ApJ...791...93A}. Figures~\ref{fig:C_AMS} and \ref{fig:O_AMS} show the calculated spectra compared with ACE/CRIS and AMS-02 data. The calculated spectra are integrated over the time period corresponding to the AMS-02 data taking (from May 2011 to May 2016), where the quoted error bars include both statistical and systematic uncertainties.

\begin{figure}[!tb]
\centerline{
 \includegraphics[width=0.49\textwidth]{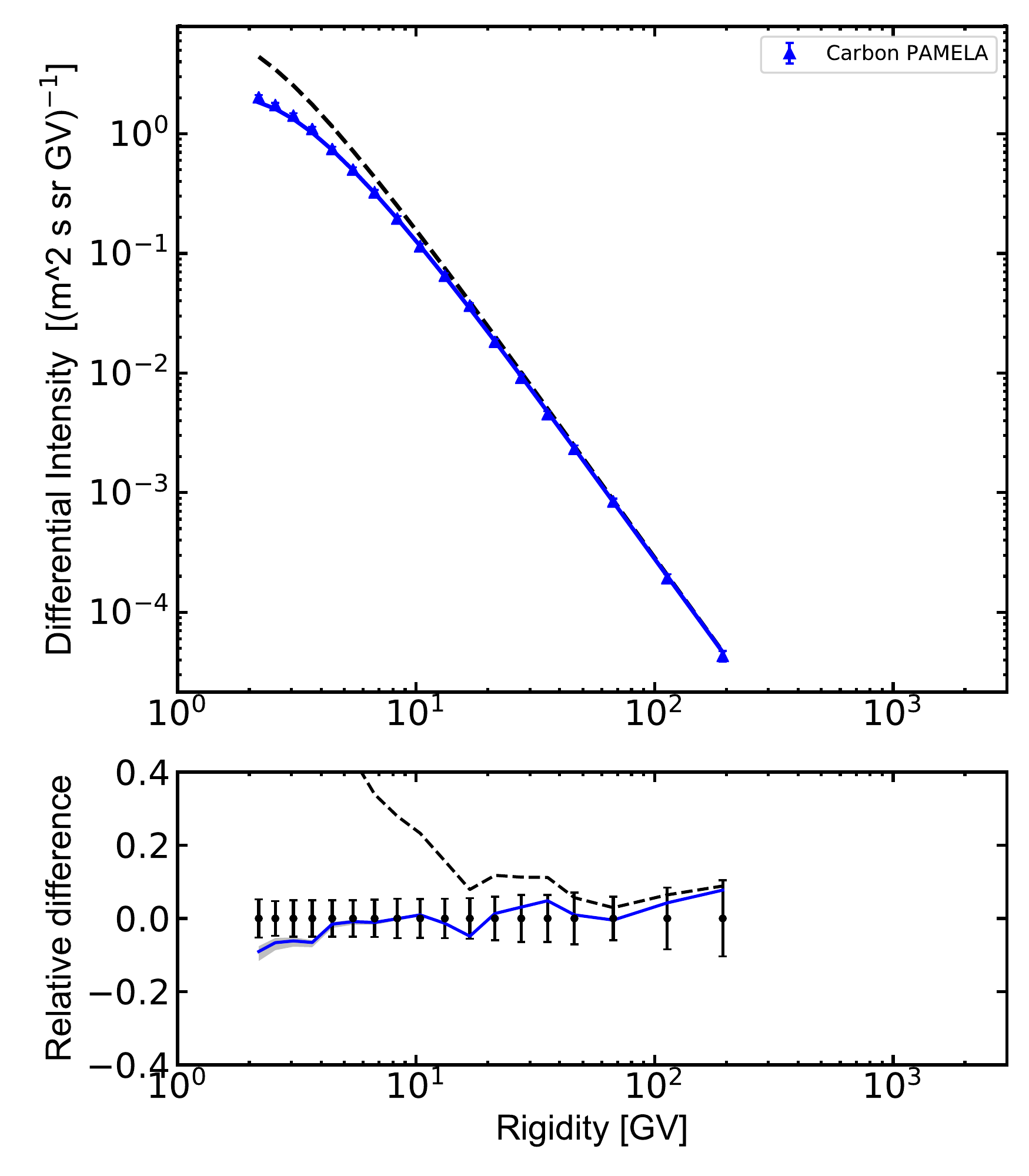}
 }
 \caption{Differential intensity of carbon from PAMELA experiment. Line coding is the same as in Figure~\ref{fig:C_AMS}. The \galprop{} LIS is multiplied by 0.85 (dashed line) to match the experimental data.
}
 \label{fig:C_PAMELA}
\end{figure}

\begin{figure*}[!tb]
 \includegraphics[width=0.49\textwidth]{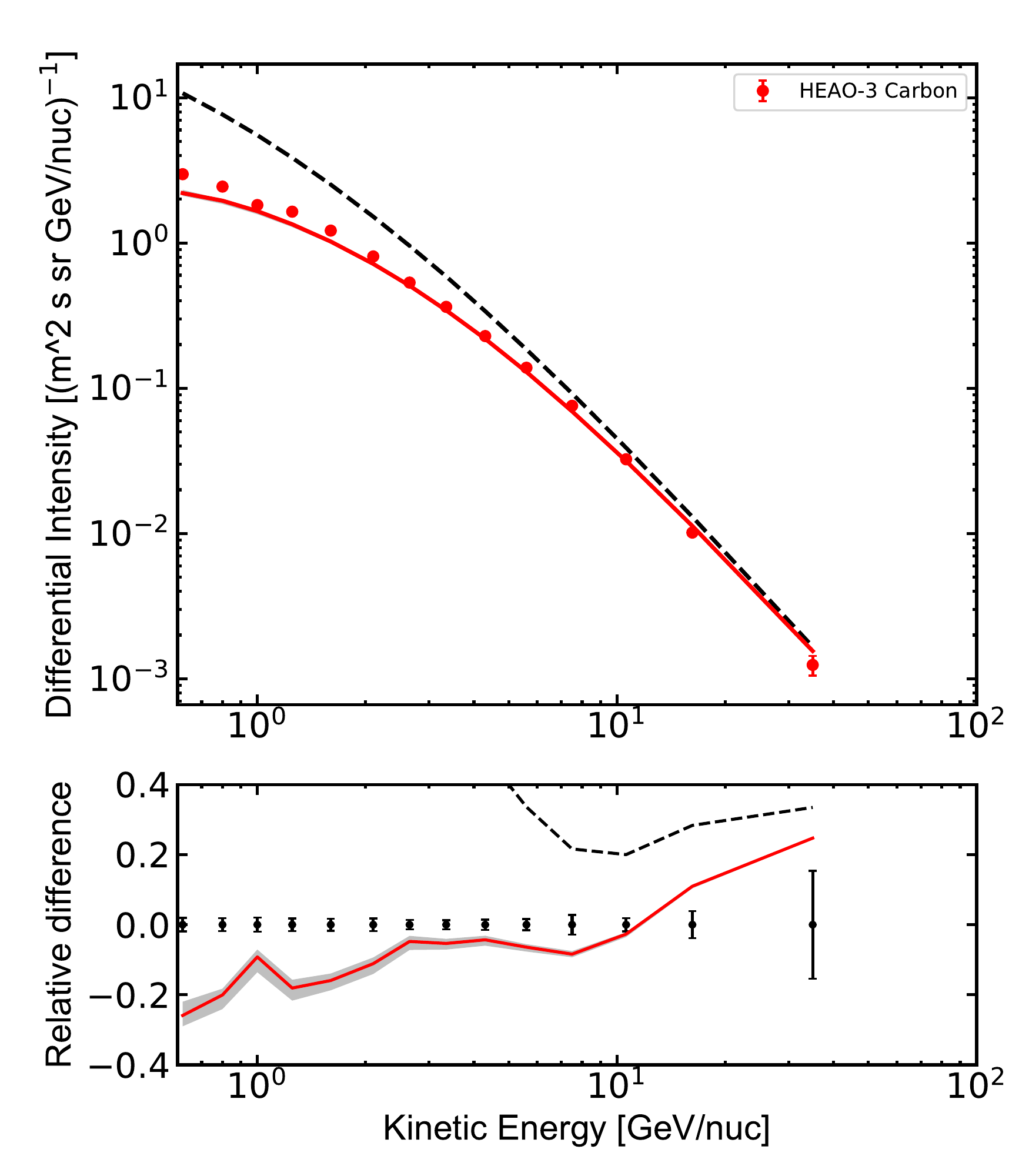}\hfill
 \includegraphics[width=0.49\textwidth]{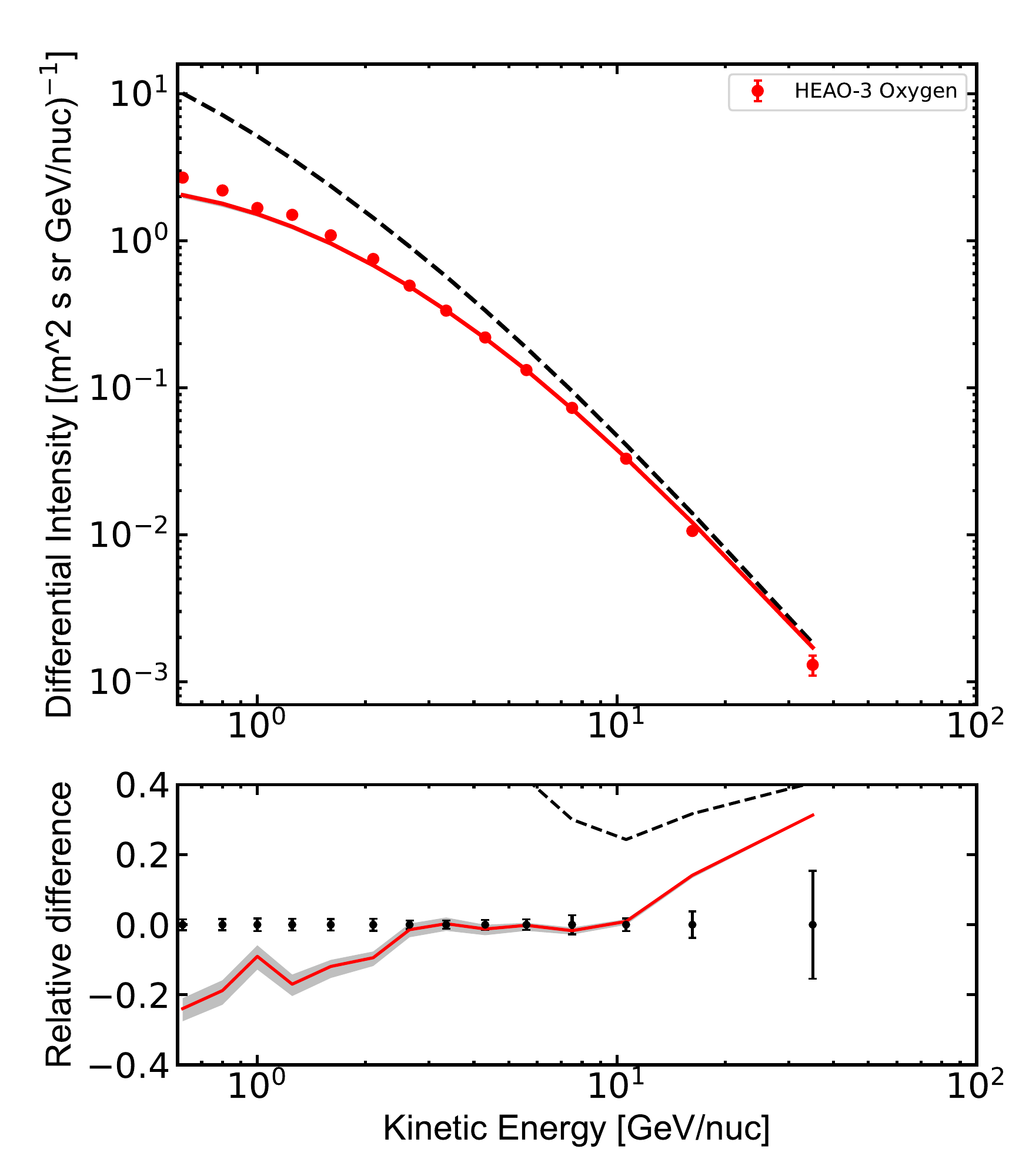}
 \caption{Differential intensity of carbon from HEAO-3 experiment. Line coding is the same as in Figure~\ref{fig:C_AMS}. 
}
 \label{fig:CO_HEAO3}
\end{figure*}

In Figure~\ref{fig:C_PAMELA} the modulated spectrum is compared with the PAMELA data taken during the period of the solar minimum spanning from July 2006 to March 2008. \imos{The \galprop{} LIS is multiplied by 0.85 to match the experimental data. We do not speculate about the origin of this factor, but point out that it may be connected with the evaluation of the selection efficiency \citep[see footnote 24 in][]{2013ApJ...765...91A}.} Figure~\ref{fig:CO_HEAO3} shows a comparison with HEAO-3 data collected from Oct 1979 to June 1980. To reproduce the proper time span for each experiment, the \helmod{} modulated spectra were evaluated for each Carrington rotation during the period of observations, and then the results were used to evaluate a unique normalized probability function for the modulation tool as described in Section 3.1 of \citet{2017ApJ...840..115B}.

Note that the AMS-02 measurements of the carbon spectrum \citep{2017PhRvL.119} are distinctly different from the results of the previous experiments, which show 20--25\% lower intensity above 20 GV (see Figure~\ref{fig:HighEnergy_He_C_O}). The discrepancy is even larger for oxygen (same Figure). Therefore, in this paper, for each presented dataset, a normalization factor that rescales the LIS to the published values is calculated, while AMS-02 data are used for the described MCMC procedure due to the small systematic and statistical uncertainties. The renormalized spectra generally agree well with the available data, but some discrepancies with earlier experiments remain (Figure~\ref{fig:CO_HEAO3}). 

\section{High Energy spectrum} \label{Sect::highenergy}

\begin{figure*}[!tb]
 \includegraphics[width=0.49\textwidth]{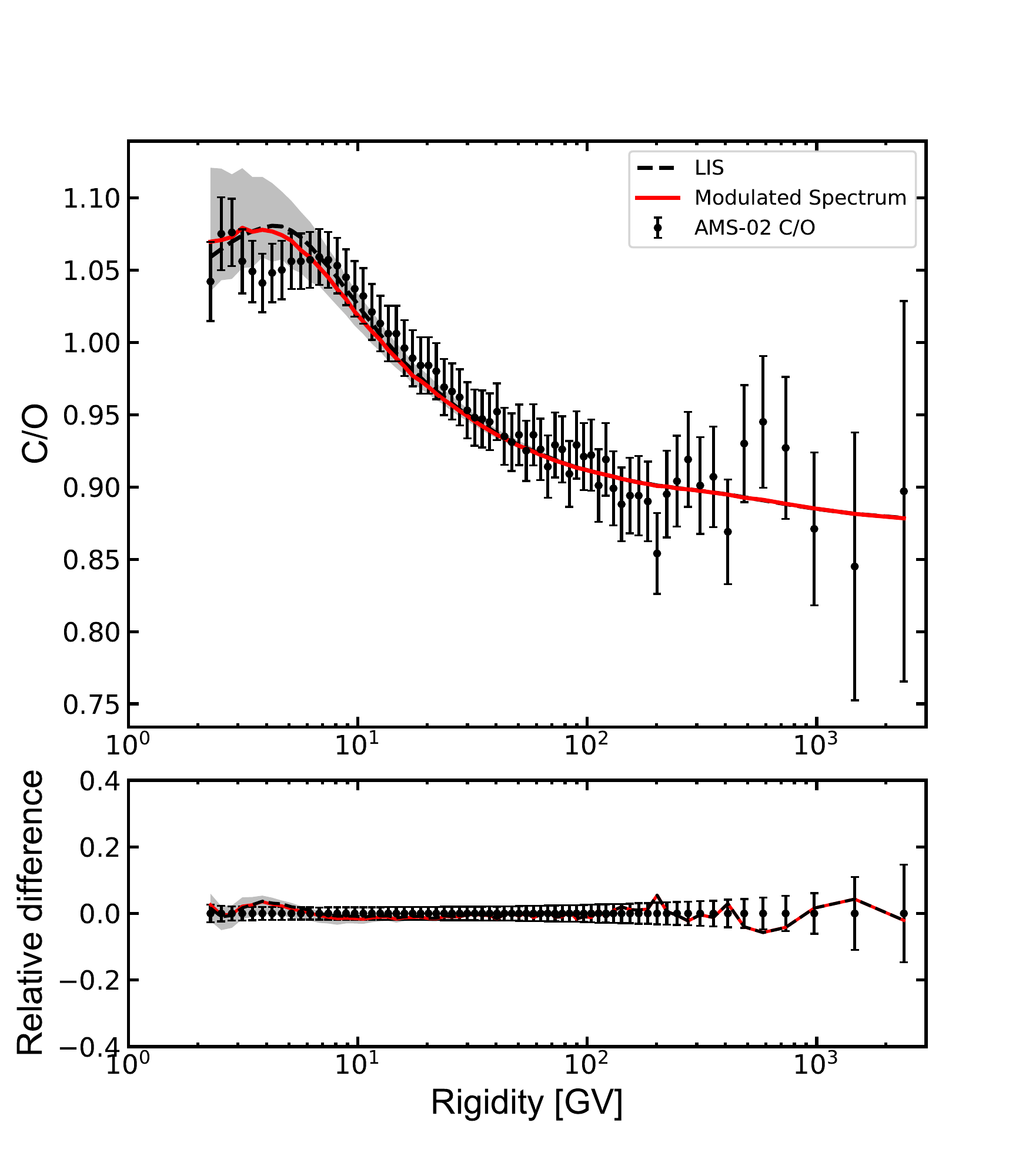}\hfill
 \includegraphics[width=0.49\textwidth]{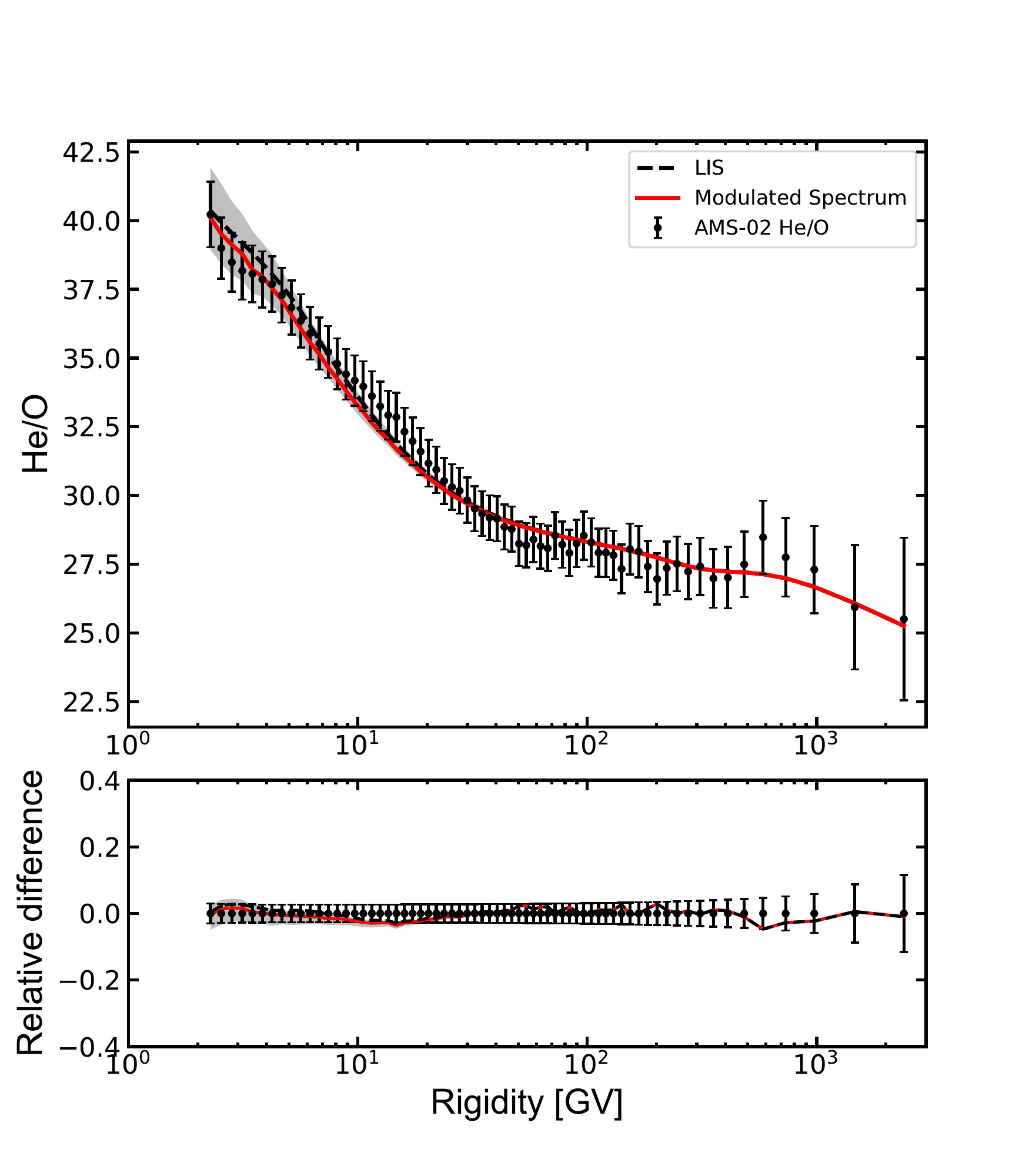}
 \caption{The carbon/oxygen (left) and helium/oxygen (right) ratios as measured by AMS-02 are compared with calculated ratios for LIS (dashed line) and modulated ratios (red line).
 \label{fig:CoO}}
\end{figure*}

\begin{figure*}[!tb]
\centerline{
\includegraphics[width=0.8\textwidth]{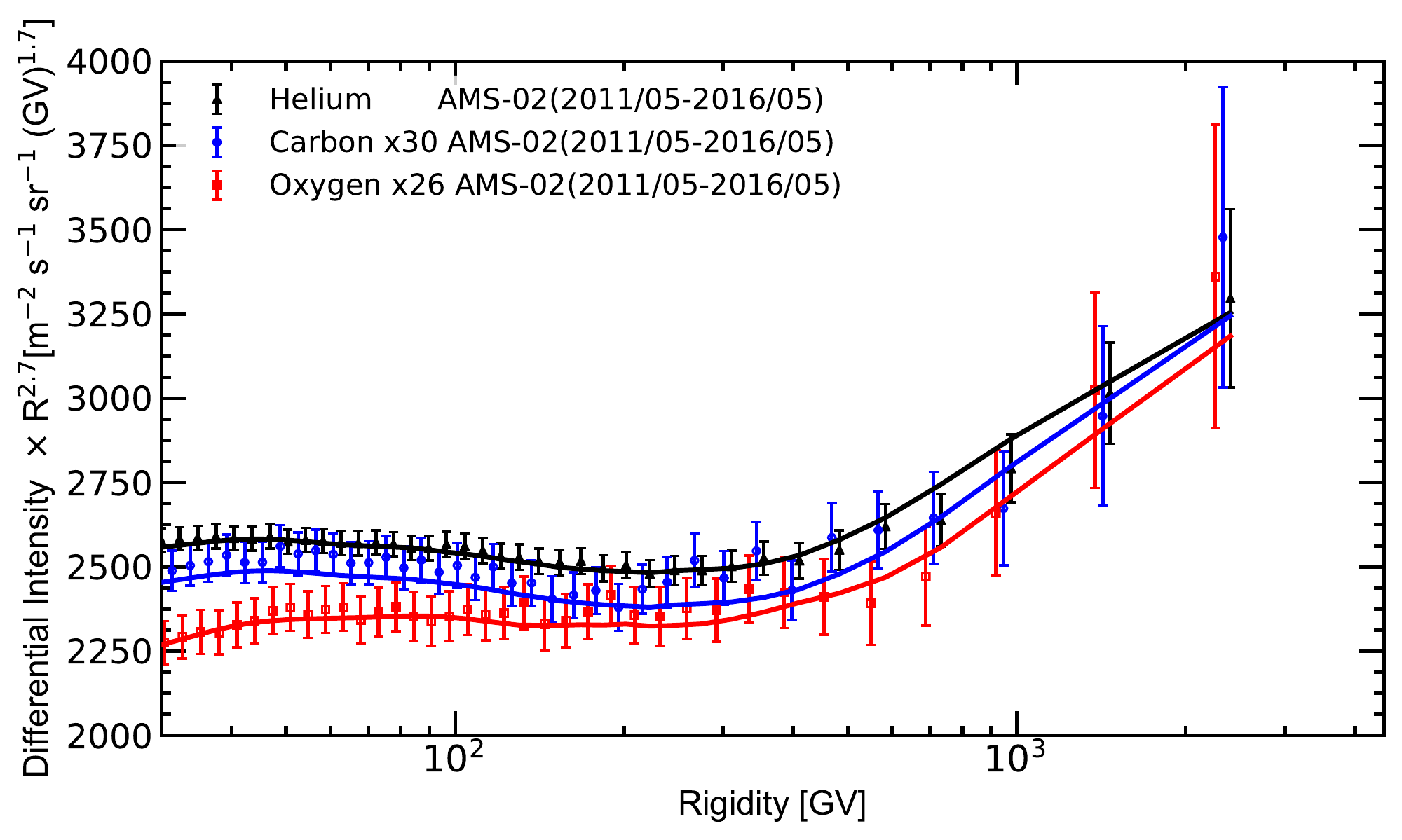}
}
\caption{The AMS-02 measurements of helium (black triangles), carbon (blue points), and oxygen (red squares)
are compared with the corresponding \galprop{}-LIS modulated with \helmod{} (solid lines).
\label{fig:HighEnergy_HeCO_Same}}
\end{figure*}

Direct measurements of primary CR species and detection of their spectral features may be able to provide a hint at the origin and properties of CR acceleration sites and/or properties of the interstellar medium. Unexpected flattening and breaks in the spectra of CR protons, helium, and heavier nuclei observed by ATIC \citep{2009BRASP..73..564P}, CREAM \citep{2010ApJ...714L..89A,2011ApJ...728..122Y}, PAMELA \citep{2011Sci...332...69A}, and more recently by AMS-02 \citep{2015PhRvL.114q1103A,2015PhRvL.115u1101A,2017PhRvL.119}, that can be seen in Figure~\ref{fig:HighEnergy_He_C_O}, stimulated a rich discussion of the origin of high-energy hardening in the spectra of CR species \citep[see, e.g.,][]{2012ApJ...752...68V,2013A&A...555A..48B,2015ApJ...815L...1T,2016PhRvD..93h3001O}. AMS-02 measurements indicate that the values of the spectral indices of carbon and oxygen with high precision resemble the value of the helium index (see Figure~\ref{fig:CoO} and \ref{fig:HighEnergy_HeCO_Same}). This supports the idea that the primary species are likely accelerated in the same processes or/and sources. In this scenario the break at high energies may be the intrinsic property of the sources or of the acceleration mechanism itself, thus requiring an \textit{ad hoc} break in the injection spectra.

\begin{deluxetable*}{r@{\hskip 4 pt}l@{\hskip 10 pt}r@{\hskip 4 pt}l@{\hskip 10 pt}r@{\hskip 4 pt}l|r@{\hskip 4 pt}l@{\hskip 10 pt}r@{\hskip 4 pt}l@{\hskip 10 pt}r@{\hskip 4 pt}l}[!tb]
\tablecolumns{12}
\tablewidth{0pt}
\tablecaption{Parameters of the analytical fits to the carbon and oxygen LIS\label{tbl-4}}
\tablehead{
\multicolumn{6}{c}{Carbon, $R_d=2$ GV} &
\multicolumn{6}{c}{Oxygen, $R_d=2$ GV} 
}
\startdata
$a_0=$   & 0.33318   &$p_{-2}=$ & 6.864                &$g_1=$& 0      & $a_0=$ & 0       &$p_{-2}=$& 0         &$g_1=$& 6910    \\
$a_1=$   & --28.356   &$p_{-1}=$ & --16.169              &$h_1=$& 0      & $a_1=$ & 0       &$p_{-1}=$& 0         &$h_1=$& 134.86  \\
$a_2=$   & 40.530    &$p_0=$    & 40.68                &$l_1=$& 0      & $a_2=$ & --8.7059 &$p_0=$   & 146.19    &$l_1=$&1.3348   \\
$a_3=$   & --8.8178   &$p_1=$    & --0.0098494           &$g_2=$& 0      & $a_3=$ & 28.64   &$p_1=$   & 0.0023718 &$g_2=$& --10698.7 \\
$a_4=$   & 0         &$p_2=$    & 1.992$\times 10^{-7}$ &$h_2=$& 0      & $a_4=$ & --13.92  &$p_2=$   & 0         &$h_2=$&130.448  \\
$a_5=$   & 0         &$f=$      & 1.7727               &$l_2=$& 0      & $a_5=$ & 2.03539 &$f=$     & 0         &$l_2=$&0.0879   \\
$b=$     & 686.15    &$g_0=$    & 4989                 &$m=$& --515.22   & $b=$   & 0       &$g_0=$   & --224.146  &$m=$&0          \\
$c=$     & 31.045    &$h_0=$    & 86.075               &$n=$& 5.2815    & $c=$   & 0       &$h_0=$   & 0.920357  &$n=$&0           \\
$d=$     & 127.83    &$l_0=$    & 1                    &$o=$&1         & $d=$   & 0       &$l_0=$   & 1         &$o=$&0   
\enddata
\end{deluxetable*}

Another reasonable hypothesis is to assume a change in the slope of the diffusion coefficient around 350 GV \citep{2012ApJ...752...68V} that would affect all CR species whose spectra and breaks will be aligned automatically. As discussed by \citet{2017ApJ...840..115B}, the flattening of helium, carbon, and oxygen spectra is then reproduced if the index of the rigidity dependence of the diffusion coefficient $\delta$ is reduced above the break rigidity by $\Delta\delta\approx0.27$. In the framework of the model discussed in this paper that means changing $\delta$ from 0.415 to $\sim$0.15. However, the discrimination of the two scenarios cannot be made without accurate measurements of secondary nuclei. A forthcoming paper will report on the analysis of the secondary species, which is currently underway. This subsequent work takes advantage of the just-published AMS-02 observations of CR lithium, beryllium, and boron \citep{PhysRevLett.120-2018}.

In addition to the plots and the tabulated data presented in Section \ref{Sect::DataAtEarth} and Appendix~\ref{app:SupMat}, the analytical expression is provided for the GALPROP LIS, from 3 MV up to 10 TV: 
\begin{align}\label{EQ::an}
&F(R)\times R^{2.7} = \\
&\left\{
\begin{array}{ll}
\sum_{i=0}^5 a_i R^{i} + \frac{b R}{c+d R^2},       &R\le R_d,\smallskip\\ \nonumber
\sum_{i=-2}^2 p_i R^{i} \! + \! f\sqrt R \! + \! \sum_{i=0}^2 \frac{g_i}{h_i+l_i R} \! + \! \frac{m}{n+o R^2} ,&R>R_d, \nonumber
\end{array}
\right.\nonumber
\end{align}
where $a_i, b, c, d, e, f, g_i, h_i, l_i, m, n, o, p_i$ are the numerical coefficients summarized in Table~\ref{tbl-4}, and $R$ is the rigidity in GV. \imos{The fit is tuned to agree with Voyager 1 measurements at low rigidities and matches the \galprop-calculated LIS in the energy range where no data points are available.} The derived expressions are virtually identical, within 1\%--5\%, to numerical solutions spanning over 5 orders of magnitude in rigidity, including the spectral flattening at high energies. The search for the analytic solutions -- using the same algorithm of \citet{2017ApJ...840..115B} -- was guided by the advanced MCMC fitting procedure such as Eureqa\footnote{http://www.nutonian.com/products/eureqa/}.

\section{Summary}

The helium, carbon, and oxygen LIS derived in the current work provide a good description of the Voyager 1, ACE/CRIS, HEAO-3, PAMELA, and AMS-02 data over the energy range from MeV/n to tens of TeV/n. The work presented in this paper demonstrates that the CR data collected during the  solar cycles 23 and 24 can be successfully reproduced within a single framework. This includes a fully realistic and exhaustive description of the relevant CR physics. Given their high precision, recent AMS-02 data can be used to put useful constraints of processes of particle acceleration, CR sources, properties of the interstellar medium, search for signatures of dark matter, and many others. This work complements earlier results on the proton, He, antiproton, and electron LIS illustrating a significant potential of the combined \galprop{}-\helmod{} framework.

\acknowledgements
Special thanks to Pavol Bobik, Giuliano Boella, Karel Kudela, Marian Putis, and Mario Zannoni for their continuous support of the \helmod{} project and many useful suggestions. This work is supported by ASI (Agenzia Spaziale Italiana) through a contract ASI-INFN I/002/13/0 and by ESA (European Space Agency) through a contract 4000116146/16/NL/HK. Igor Moskalenko and Troy Porter acknowledge support from NASA Grant No.~NNX17AB48G. We thank the ACE/CRIS instrument team and the ACE Science Center for providing the ACE data. The \helmod{} team acknowledges use of the OMNI data and OMNIWeb (CDAWeb, ftp) services provided by the NASA/GSFC Space Physics Data Facility\footnote{https://omniweb.gsfc.nasa.gov}.

\bibliography{bibliography}

\eject
\clearpage

\appendix
\section*{Supplementary Material}\label{app:SupMat}

\newpage

\begin{deluxetable*}{cccccccccc}[p]
\tablecolumns{10}
\tablewidth{0mm}
\tablecaption{Carbon LIS\label{Tbl-CarbonLIS}}
\tablehead{
\colhead{Rigidity} & \colhead{Differential} &
\colhead{Rigidity} & \colhead{Differential} &
\colhead{Rigidity} & \colhead{Differential} &
\colhead{Rigidity} & \colhead{Differential} &
\colhead{Rigidity} & \colhead{Differential}
\\
\colhead{GV} & \colhead{Intensity\tablenotemark{a}} &
\colhead{GV} & \colhead{Intensity\tablenotemark{a}} &
\colhead{GV} & \colhead{Intensity\tablenotemark{a}} &
\colhead{GV} & \colhead{Intensity\tablenotemark{a}} &
\colhead{GV} & \colhead{Intensity\tablenotemark{a}} 
}
\startdata
\startdata
\startdata
9.354e-02 & 1.412e-01 & 5.922e-01 & 5.255e+00 & 4.965e+00 & 1.048e+00 & 1.425e+02 & 1.262e-04 & 5.518e+03 & 9.869e-09\\
9.678e-02 & 1.526e-01 & 6.131e-01 & 5.489e+00 & 5.218e+00 & 9.316e-01 & 1.523e+02 & 1.049e-04 & 5.906e+03 & 8.318e-09\\
1.001e-01 & 1.647e-01 & 6.348e-01 & 5.724e+00 & 5.487e+00 & 8.262e-01 & 1.629e+02 & 8.720e-05 & 6.321e+03 & 7.010e-09\\
1.036e-01 & 1.778e-01 & 6.573e-01 & 5.959e+00 & 5.773e+00 & 7.308e-01 & 1.742e+02 & 7.251e-05 & 6.765e+03 & 5.908e-09\\
1.072e-01 & 1.919e-01 & 6.806e-01 & 6.192e+00 & 6.077e+00 & 6.450e-01 & 1.863e+02 & 6.032e-05 & 7.241e+03 & 4.979e-09\\
1.109e-01 & 2.071e-01 & 7.048e-01 & 6.420e+00 & 6.402e+00 & 5.679e-01 & 1.993e+02 & 5.019e-05 & 7.750e+03 & 4.196e-09\\
1.147e-01 & 2.235e-01 & 7.299e-01 & 6.641e+00 & 6.747e+00 & 4.988e-01 & 2.132e+02 & 4.177e-05 & 8.295e+03 & 3.536e-09\\
1.187e-01 & 2.411e-01 & 7.560e-01 & 6.854e+00 & 7.115e+00 & 4.371e-01 & 2.280e+02 & 3.479e-05 & 8.879e+03 & 2.980e-09\\
1.228e-01 & 2.601e-01 & 7.830e-01 & 7.056e+00 & 7.507e+00 & 3.821e-01 & 2.439e+02 & 2.898e-05 & 9.503e+03 & 2.511e-09\\
1.270e-01 & 2.806e-01 & 8.111e-01 & 7.246e+00 & 7.926e+00 & 3.332e-01 & 2.610e+02 & 2.416e-05 & 1.017e+04 & 2.116e-09\\
1.314e-01 & 3.027e-01 & 8.402e-01 & 7.422e+00 & 8.372e+00 & 2.898e-01 & 2.792e+02 & 2.014e-05 & 1.089e+04 & 1.783e-09\\
1.360e-01 & 3.264e-01 & 8.705e-01 & 7.583e+00 & 8.849e+00 & 2.514e-01 & 2.987e+02 & 1.681e-05 & 1.165e+04 & 1.502e-09\\
1.407e-01 & 3.519e-01 & 9.019e-01 & 7.726e+00 & 9.357e+00 & 2.176e-01 & 3.195e+02 & 1.403e-05 & 1.247e+04 & 1.266e-09\\
1.456e-01 & 3.794e-01 & 9.346e-01 & 7.849e+00 & 9.900e+00 & 1.880e-01 & 3.419e+02 & 1.172e-05 & 1.335e+04 & 1.066e-09\\
1.506e-01 & 4.089e-01 & 9.685e-01 & 7.953e+00 & 1.048e+01 & 1.620e-01 & 3.658e+02 & 9.800e-06 & 1.429e+04 & 8.986e-10\\
1.558e-01 & 4.407e-01 & 1.004e+00 & 8.036e+00 & 1.110e+01 & 1.394e-01 & 3.914e+02 & 8.197e-06 & 1.529e+04 & 7.571e-10\\
1.612e-01 & 4.748e-01 & 1.041e+00 & 8.098e+00 & 1.176e+01 & 1.197e-01 & 4.187e+02 & 6.860e-06 & 1.637e+04 & 6.378e-10\\
1.668e-01 & 5.114e-01 & 1.079e+00 & 8.137e+00 & 1.247e+01 & 1.025e-01 & 4.480e+02 & 5.745e-06 & 1.752e+04 & 5.374e-10\\
1.726e-01 & 5.506e-01 & 1.119e+00 & 8.154e+00 & 1.322e+01 & 8.764e-02 & 4.794e+02 & 4.815e-06 & 1.875e+04 & 4.527e-10\\
1.785e-01 & 5.928e-01 & 1.160e+00 & 8.148e+00 & 1.403e+01 & 7.479e-02 & 5.130e+02 & 4.037e-06 & 2.007e+04 & 3.814e-10\\
1.847e-01 & 6.379e-01 & 1.203e+00 & 8.120e+00 & 1.489e+01 & 6.372e-02 & 5.489e+02 & 3.387e-06 & 2.148e+04 & 3.213e-10\\
1.911e-01 & 6.863e-01 & 1.248e+00 & 8.070e+00 & 1.582e+01 & 5.419e-02 & 5.874e+02 & 2.843e-06 & 2.299e+04 & 2.707e-10\\
1.978e-01 & 7.382e-01 & 1.295e+00 & 7.998e+00 & 1.680e+01 & 4.602e-02 & 6.286e+02 & 2.387e-06 & 2.461e+04 & 2.280e-10\\
2.046e-01 & 7.936e-01 & 1.343e+00 & 7.904e+00 & 1.786e+01 & 3.903e-02 & 6.727e+02 & 2.006e-06 & 2.634e+04 & 1.921e-10\\
2.117e-01 & 8.530e-01 & 1.394e+00 & 7.789e+00 & 1.899e+01 & 3.305e-02 & 7.198e+02 & 1.686e-06 & 2.819e+04 & 1.618e-10\\
2.190e-01 & 9.164e-01 & 1.448e+00 & 7.653e+00 & 2.020e+01 & 2.794e-02 & 7.703e+02 & 1.417e-06 & 3.017e+04 & 1.363e-10\\
2.266e-01 & 9.841e-01 & 1.503e+00 & 7.497e+00 & 2.149e+01 & 2.360e-02 & 8.243e+02 & 1.192e-06 & 3.230e+04 & 1.148e-10\\
2.345e-01 & 1.056e+00 & 1.561e+00 & 7.321e+00 & 2.287e+01 & 1.990e-02 & 8.822e+02 & 1.003e-06 & 3.457e+04 & 9.671e-11\\
2.426e-01 & 1.133e+00 & 1.622e+00 & 7.127e+00 & 2.435e+01 & 1.676e-02 & 9.441e+02 & 8.441e-07 & 3.700e+04 & 8.146e-11\\
2.511e-01 & 1.215e+00 & 1.685e+00 & 6.916e+00 & 2.593e+01 & 1.410e-02 & 1.010e+03 & 7.105e-07 & 3.960e+04 & 6.862e-11\\
2.598e-01 & 1.303e+00 & 1.751e+00 & 6.689e+00 & 2.762e+01 & 1.185e-02 & 1.081e+03 & 5.983e-07 & 4.239e+04 & 5.780e-11\\
2.688e-01 & 1.395e+00 & 1.821e+00 & 6.448e+00 & 2.943e+01 & 9.947e-03 & 1.157e+03 & 5.039e-07 & 4.537e+04 & 4.868e-11\\
2.781e-01 & 1.494e+00 & 1.893e+00 & 6.194e+00 & 3.137e+01 & 8.343e-03 & 1.238e+03 & 4.245e-07 & 4.856e+04 & 4.100e-11\\
2.878e-01 & 1.599e+00 & 1.969e+00 & 5.927e+00 & 3.344e+01 & 6.991e-03 & 1.325e+03 & 3.576e-07 & 5.197e+04 & 3.453e-11\\
2.978e-01 & 1.709e+00 & 2.049e+00 & 5.651e+00 & 3.566e+01 & 5.854e-03 & 1.418e+03 & 3.013e-07 & 5.563e+04 & 2.909e-11\\
3.081e-01 & 1.826e+00 & 2.133e+00 & 5.367e+00 & 3.803e+01 & 4.898e-03 & 1.518e+03 & 2.539e-07 & 5.954e+04 & 2.450e-11\\
3.189e-01 & 1.949e+00 & 2.221e+00 & 5.079e+00 & 4.057e+01 & 4.096e-03 & 1.625e+03 & 2.139e-07 & 6.373e+04 & 2.063e-11\\
3.300e-01 & 2.079e+00 & 2.314e+00 & 4.788e+00 & 4.329e+01 & 3.422e-03 & 1.739e+03 & 1.803e-07 & 6.821e+04 & 1.738e-11\\
3.414e-01 & 2.215e+00 & 2.411e+00 & 4.498e+00 & 4.620e+01 & 2.858e-03 & 1.861e+03 & 1.520e-07 & 7.301e+04 & 1.464e-11\\
3.533e-01 & 2.358e+00 & 2.514e+00 & 4.210e+00 & 4.932e+01 & 2.385e-03 & 1.992e+03 & 1.281e-07 & 7.814e+04 & 1.233e-11\\
3.656e-01 & 2.507e+00 & 2.622e+00 & 3.926e+00 & 5.265e+01 & 1.989e-03 & 2.132e+03 & 1.080e-07 & 8.364e+04 & 1.038e-11\\
3.784e-01 & 2.663e+00 & 2.735e+00 & 3.649e+00 & 5.621e+01 & 1.659e-03 & 2.282e+03 & 9.100e-08 & 8.952e+04 & 8.743e-12\\
3.916e-01 & 2.826e+00 & 2.855e+00 & 3.380e+00 & 6.003e+01 & 1.382e-03 & 2.442e+03 & 7.671e-08 & 9.582e+04 & 7.363e-12\\
4.053e-01 & 2.995e+00 & 2.981e+00 & 3.120e+00 & 6.412e+01 & 1.151e-03 & 2.613e+03 & 6.467e-08 & 1.026e+05 & 6.201e-12\\
4.194e-01 & 3.170e+00 & 3.115e+00 & 2.871e+00 & 6.849e+01 & 9.589e-04 & 2.797e+03 & 5.451e-08 & 1.098e+05 & 5.222e-12\\
4.341e-01 & 3.352e+00 & 3.256e+00 & 2.633e+00 & 7.317e+01 & 7.982e-04 & 2.994e+03 & 4.595e-08 & 1.175e+05 & 4.398e-12\\
4.493e-01 & 3.539e+00 & 3.405e+00 & 2.407e+00 & 7.817e+01 & 6.643e-04 & 3.204e+03 & 3.874e-08 & 1.258e+05 & 3.703e-12\\
4.650e-01 & 3.733e+00 & 3.563e+00 & 2.193e+00 & 8.353e+01 & 5.527e-04 & 3.429e+03 & 3.265e-08 & 1.346e+05 & 3.119e-12\\
4.813e-01 & 3.934e+00 & 3.730e+00 & 1.992e+00 & 8.927e+01 & 4.597e-04 & 3.670e+03 & 2.752e-08 & 1.441e+05 & 2.626e-12\\
4.982e-01 & 4.140e+00 & 3.906e+00 & 1.804e+00 & 9.541e+01 & 3.823e-04 & 3.928e+03 & 2.320e-08 & 1.542e+05 & 2.212e-12\\
5.156e-01 & 4.353e+00 & 4.094e+00 & 1.628e+00 & 1.020e+02 & 3.179e-04 & 4.205e+03 & 1.956e-08 & 1.650e+05 & 1.862e-12\\
5.338e-01 & 4.571e+00 & 4.293e+00 & 1.465e+00 & 1.090e+02 & 2.643e-04 & 4.500e+03 & 1.648e-08 & 1.767e+05 & 1.569e-12\\
5.526e-01 & 4.795e+00 & 4.503e+00 & 1.314e+00 & 1.165e+02 & 2.197e-04 & 4.817e+03 & 1.389e-08 & 1.891e+05 & 1.317e-12\\
5.720e-01 & 5.023e+00 & 4.727e+00 & 1.175e+00 & 1.246e+02 & 1.826e-04 & 5.155e+03 & 1.171e-08 & 2.024e+05 & 1.054e-12
\enddata

\enddata
\enddata
\tablenotetext{a}{Differential Intensity units: (m$^2$ s sr GV)$^{-1}$.}
\end{deluxetable*}

\begin{deluxetable*}{cccccccccc}[p]
\tablecolumns{10}
\tablewidth{0mm}
\tablecaption{Oxygen LIS\label{Tbl-OxygenLIS}}
\tablehead{
\colhead{Rigidity} & \colhead{Differential} &
\colhead{Rigidity} & \colhead{Differential} &
\colhead{Rigidity} & \colhead{Differential} &
\colhead{Rigidity} & \colhead{Differential} &
\colhead{Rigidity} & \colhead{Differential}
\\
\colhead{GV} & \colhead{Intensity\tablenotemark{a}} &
\colhead{GV} & \colhead{Intensity\tablenotemark{a}} &
\colhead{GV} & \colhead{Intensity\tablenotemark{a}} &
\colhead{GV} & \colhead{Intensity\tablenotemark{a}} &
\colhead{GV} & \colhead{Intensity\tablenotemark{a}} 
}
\startdata
\startdata
9.714e-02 & 1.655e-01 & 6.150e-01 & 5.553e+00 & 5.156e+00 & 9.070e-01 & 1.479e+02 & 1.266e-04 & 5.730e+03 & 1.025e-08\\
1.005e-01 & 1.784e-01 & 6.367e-01 & 5.784e+00 & 5.419e+00 & 8.073e-01 & 1.582e+02 & 1.054e-04 & 6.133e+03 & 8.641e-09\\
1.040e-01 & 1.922e-01 & 6.592e-01 & 6.014e+00 & 5.698e+00 & 7.172e-01 & 1.692e+02 & 8.768e-05 & 6.564e+03 & 7.281e-09\\
1.076e-01 & 2.071e-01 & 6.826e-01 & 6.241e+00 & 5.995e+00 & 6.359e-01 & 1.809e+02 & 7.298e-05 & 7.026e+03 & 6.135e-09\\
1.113e-01 & 2.231e-01 & 7.068e-01 & 6.463e+00 & 6.311e+00 & 5.626e-01 & 1.935e+02 & 6.076e-05 & 7.520e+03 & 5.169e-09\\
1.151e-01 & 2.404e-01 & 7.319e-01 & 6.680e+00 & 6.648e+00 & 4.968e-01 & 2.070e+02 & 5.060e-05 & 8.048e+03 & 4.355e-09\\
1.191e-01 & 2.589e-01 & 7.580e-01 & 6.889e+00 & 7.006e+00 & 4.377e-01 & 2.214e+02 & 4.216e-05 & 8.614e+03 & 3.669e-09\\
1.233e-01 & 2.789e-01 & 7.850e-01 & 7.087e+00 & 7.389e+00 & 3.848e-01 & 2.368e+02 & 3.513e-05 & 9.220e+03 & 3.091e-09\\
1.275e-01 & 3.003e-01 & 8.131e-01 & 7.274e+00 & 7.796e+00 & 3.375e-01 & 2.533e+02 & 2.929e-05 & 9.868e+03 & 2.604e-09\\
1.319e-01 & 3.234e-01 & 8.423e-01 & 7.448e+00 & 8.231e+00 & 2.953e-01 & 2.710e+02 & 2.444e-05 & 1.056e+04 & 2.194e-09\\
1.365e-01 & 3.481e-01 & 8.725e-01 & 7.606e+00 & 8.694e+00 & 2.578e-01 & 2.899e+02 & 2.040e-05 & 1.131e+04 & 1.848e-09\\
1.412e-01 & 3.747e-01 & 9.039e-01 & 7.748e+00 & 9.189e+00 & 2.246e-01 & 3.102e+02 & 1.703e-05 & 1.210e+04 & 1.556e-09\\
1.461e-01 & 4.033e-01 & 9.366e-01 & 7.871e+00 & 9.717e+00 & 1.953e-01 & 3.318e+02 & 1.423e-05 & 1.295e+04 & 1.311e-09\\
1.512e-01 & 4.339e-01 & 9.705e-01 & 7.975e+00 & 1.028e+01 & 1.694e-01 & 3.550e+02 & 1.190e-05 & 1.386e+04 & 1.104e-09\\
1.564e-01 & 4.668e-01 & 1.006e+00 & 8.057e+00 & 1.088e+01 & 1.467e-01 & 3.798e+02 & 9.959e-06 & 1.484e+04 & 9.298e-10\\
1.618e-01 & 5.020e-01 & 1.042e+00 & 8.118e+00 & 1.153e+01 & 1.267e-01 & 4.064e+02 & 8.338e-06 & 1.588e+04 & 7.831e-10\\
1.674e-01 & 5.398e-01 & 1.081e+00 & 8.156e+00 & 1.221e+01 & 1.092e-01 & 4.348e+02 & 6.986e-06 & 1.700e+04 & 6.595e-10\\
1.732e-01 & 5.803e-01 & 1.120e+00 & 8.169e+00 & 1.295e+01 & 9.398e-02 & 4.653e+02 & 5.857e-06 & 1.819e+04 & 5.553e-10\\
1.792e-01 & 6.237e-01 & 1.162e+00 & 8.159e+00 & 1.373e+01 & 8.070e-02 & 4.979e+02 & 4.913e-06 & 1.947e+04 & 4.676e-10\\
1.854e-01 & 6.700e-01 & 1.205e+00 & 8.124e+00 & 1.457e+01 & 6.916e-02 & 5.327e+02 & 4.124e-06 & 2.084e+04 & 3.938e-10\\
1.918e-01 & 7.196e-01 & 1.249e+00 & 8.063e+00 & 1.547e+01 & 5.916e-02 & 5.701e+02 & 3.463e-06 & 2.231e+04 & 3.315e-10\\
1.985e-01 & 7.727e-01 & 1.296e+00 & 7.979e+00 & 1.643e+01 & 5.051e-02 & 6.100e+02 & 2.910e-06 & 2.388e+04 & 2.792e-10\\
2.054e-01 & 8.293e-01 & 1.344e+00 & 7.870e+00 & 1.745e+01 & 4.305e-02 & 6.528e+02 & 2.446e-06 & 2.555e+04 & 2.350e-10\\
2.125e-01 & 8.898e-01 & 1.395e+00 & 7.737e+00 & 1.855e+01 & 3.663e-02 & 6.985e+02 & 2.057e-06 & 2.735e+04 & 1.979e-10\\
2.198e-01 & 9.543e-01 & 1.448e+00 & 7.583e+00 & 1.972e+01 & 3.112e-02 & 7.475e+02 & 1.731e-06 & 2.928e+04 & 1.666e-10\\
2.275e-01 & 1.023e+00 & 1.503e+00 & 7.409e+00 & 2.097e+01 & 2.640e-02 & 7.999e+02 & 1.457e-06 & 3.133e+04 & 1.403e-10\\
2.354e-01 & 1.096e+00 & 1.561e+00 & 7.215e+00 & 2.231e+01 & 2.236e-02 & 8.561e+02 & 1.226e-06 & 3.354e+04 & 1.181e-10\\
2.435e-01 & 1.174e+00 & 1.621e+00 & 7.003e+00 & 2.375e+01 & 1.891e-02 & 9.161e+02 & 1.033e-06 & 3.590e+04 & 9.940e-11\\
2.520e-01 & 1.257e+00 & 1.684e+00 & 6.774e+00 & 2.528e+01 & 1.598e-02 & 9.804e+02 & 8.699e-07 & 3.842e+04 & 8.367e-11\\
2.607e-01 & 1.345e+00 & 1.750e+00 & 6.531e+00 & 2.693e+01 & 1.348e-02 & 1.049e+03 & 7.330e-07 & 4.112e+04 & 7.043e-11\\
2.698e-01 & 1.438e+00 & 1.819e+00 & 6.275e+00 & 2.868e+01 & 1.136e-02 & 1.123e+03 & 6.177e-07 & 4.402e+04 & 5.929e-11\\
2.791e-01 & 1.537e+00 & 1.891e+00 & 6.009e+00 & 3.056e+01 & 9.565e-03 & 1.202e+03 & 5.206e-07 & 4.711e+04 & 4.990e-11\\
2.888e-01 & 1.641e+00 & 1.966e+00 & 5.736e+00 & 3.258e+01 & 8.045e-03 & 1.286e+03 & 4.388e-07 & 5.043e+04 & 4.200e-11\\
2.989e-01 & 1.752e+00 & 2.045e+00 & 5.456e+00 & 3.473e+01 & 6.760e-03 & 1.376e+03 & 3.699e-07 & 5.397e+04 & 3.536e-11\\
3.092e-01 & 1.869e+00 & 2.128e+00 & 5.173e+00 & 3.703e+01 & 5.675e-03 & 1.473e+03 & 3.118e-07 & 5.777e+04 & 2.976e-11\\
3.200e-01 & 1.992e+00 & 2.215e+00 & 4.889e+00 & 3.950e+01 & 4.761e-03 & 1.576e+03 & 2.629e-07 & 6.183e+04 & 2.505e-11\\
3.311e-01 & 2.121e+00 & 2.307e+00 & 4.606e+00 & 4.213e+01 & 3.990e-03 & 1.687e+03 & 2.216e-07 & 6.618e+04 & 2.108e-11\\
3.426e-01 & 2.257e+00 & 2.403e+00 & 4.324e+00 & 4.496e+01 & 3.342e-03 & 1.806e+03 & 1.869e-07 & 7.083e+04 & 1.774e-11\\
3.546e-01 & 2.400e+00 & 2.504e+00 & 4.045e+00 & 4.798e+01 & 2.796e-03 & 1.933e+03 & 1.575e-07 & 7.582e+04 & 1.493e-11\\
3.669e-01 & 2.550e+00 & 2.610e+00 & 3.770e+00 & 5.121e+01 & 2.339e-03 & 2.068e+03 & 1.328e-07 & 8.115e+04 & 1.256e-11\\
3.797e-01 & 2.707e+00 & 2.722e+00 & 3.502e+00 & 5.467e+01 & 1.955e-03 & 2.214e+03 & 1.120e-07 & 8.686e+04 & 1.057e-11\\
3.929e-01 & 2.871e+00 & 2.840e+00 & 3.242e+00 & 5.838e+01 & 1.633e-03 & 2.369e+03 & 9.446e-08 & 9.296e+04 & 8.898e-12\\
4.066e-01 & 3.042e+00 & 2.965e+00 & 2.993e+00 & 6.234e+01 & 1.363e-03 & 2.536e+03 & 7.965e-08 & 9.950e+04 & 7.488e-12\\
4.208e-01 & 3.220e+00 & 3.096e+00 & 2.754e+00 & 6.658e+01 & 1.137e-03 & 2.714e+03 & 6.715e-08 & 1.065e+05 & 6.301e-12\\
4.355e-01 & 3.405e+00 & 3.235e+00 & 2.526e+00 & 7.112e+01 & 9.487e-04 & 2.905e+03 & 5.662e-08 & 1.140e+05 & 5.302e-12\\
4.508e-01 & 3.597e+00 & 3.381e+00 & 2.310e+00 & 7.598e+01 & 7.910e-04 & 3.109e+03 & 4.774e-08 & 1.220e+05 & 4.462e-12\\
4.665e-01 & 3.795e+00 & 3.536e+00 & 2.106e+00 & 8.118e+01 & 6.593e-04 & 3.327e+03 & 4.025e-08 & 1.306e+05 & 3.755e-12\\
4.829e-01 & 3.999e+00 & 3.700e+00 & 1.914e+00 & 8.675e+01 & 5.493e-04 & 3.561e+03 & 3.393e-08 & 1.398e+05 & 3.159e-12\\
4.998e-01 & 4.208e+00 & 3.873e+00 & 1.734e+00 & 9.270e+01 & 4.575e-04 & 3.812e+03 & 2.860e-08 & 1.496e+05 & 2.658e-12\\
5.173e-01 & 4.423e+00 & 4.057e+00 & 1.567e+00 & 9.908e+01 & 3.810e-04 & 4.080e+03 & 2.411e-08 & 1.601e+05 & 2.237e-12\\
5.355e-01 & 4.643e+00 & 4.251e+00 & 1.412e+00 & 1.059e+02 & 3.172e-04 & 4.366e+03 & 2.032e-08 & 1.714e+05 & 1.882e-12\\
5.543e-01 & 4.867e+00 & 4.458e+00 & 1.268e+00 & 1.132e+02 & 2.640e-04 & 4.673e+03 & 1.713e-08 & 1.834e+05 & 1.584e-12\\
5.738e-01 & 5.093e+00 & 4.677e+00 & 1.137e+00 & 1.210e+02 & 2.197e-04 & 5.002e+03 & 1.444e-08 & 1.963e+05 & 1.306e-12\\
5.940e-01 & 5.323e+00 & 4.909e+00 & 1.017e+00 & 1.294e+02 & 1.828e-04 & 5.354e+03 & 1.217e-08 & 2.102e+05 & 9.688e-13
\enddata
\enddata
\tablenotetext{a}{Differential Intensity units: (m$^2$ s sr GV)$^{-1}$.}
\end{deluxetable*}

\end{document}